\documentclass[journal,10pt,onecolumn,draftclsnofoot,]{IEEEtran}

\usepackage{graphicx}                
\usepackage{amsmath}                 
\usepackage{amsfonts}                
\usepackage{amssymb}                 
\usepackage{cite}                    
\usepackage{hyperref}                
\usepackage{tabularx}                
\usepackage{multirow}                
\usepackage{booktabs}                
\usepackage{float}                   
\usepackage{array}
\usepackage{xcolor} 
\usepackage{adjustbox} 
\usepackage{wrapfig} 

\usepackage{makecell}

\usepackage{bbding}

\usepackage[table]{xcolor} 
\usepackage{colortbl} 
\definecolor{good}{RGB}{146,208,80} 
\definecolor{moderate}{RGB}{255,235,132} 
\definecolor{poor}{RGB}{255,153,153}

\usepackage{float}

\usepackage{tabularray}
\usepackage[USenglish]{babel}
\usepackage[nodayofweek,level]{datetime}
\usepackage[utf8]{inputenc}
\usepackage{rotating}
\usepackage{array,amsmath,graphicx}

\usepackage{pdflscape}
\usepackage{afterpage}
\usepackage{capt-of}

\usepackage{lipsum}
\usepackage{cite}
\usepackage{csquotes}

\usepackage{tabu}
\def\tsc#1{\csdef{#1}{\textsc{\lowercase{#1}}\xspace}}
\tsc{WGM}
\tsc{QE}
\tsc{EP}
\tsc{PMS}
\tsc{BEC}
\tsc{DE}

\usepackage{framed}

\usepackage{tikz,lipsum,lmodern}
\usepackage[most]{tcolorbox}

\usetikzlibrary{trees,decorations,shadows}
\usetikzlibrary{arrows.meta}
\usepackage{forest}
\usetikzlibrary{external}

\let\oldtikzexternalgetnextfilename\tikzexternalgetnextfilename \renewcommand{\tikzexternalgetnextfilename}[1]{\oldtikzexternalgetnextfilename{#1}\expandafter\tikzsetnextfilename\expandafter{#1}}

\usepackage{pgfplotstable}
\usepackage[outline]{contour}
\contourlength{1.2pt}

\pgfplotsset{compat=1.13} 

\usetikzlibrary{spy}
\usetikzlibrary{calc}
\usetikzlibrary{fadings}
\usetikzlibrary{patterns}
\usetikzlibrary{shadows}
\usetikzlibrary{mindmap}
\usetikzlibrary{backgrounds}
\usetikzlibrary{shapes.symbols}
\usetikzlibrary{shapes.multipart}
\usetikzlibrary{shapes.geometric}
\usetikzlibrary{automata,positioning}
\usetikzlibrary{decorations.fractals} 
\usetikzlibrary{decorations.markings}
\usetikzlibrary{decorations.pathreplacing}
\usetikzlibrary{decorations.pathmorphing}
 
\usepackage{bm}
\usepackage{enumitem}

\usepackage{nomencl}
\makenomenclature
\tikzset{edge from parent/.style={segment angle=10,draw}}

\tikzset{
 my rounded corners/.append style={rounded corners=2pt},
}

\def\BibTeX{{\rm B\kern-.05em{\sc i\kern-.025em b}\kern-.08em
 T\kern-.1667em\lower.7ex\hbox{E}\kern-.125emX}}

\renewcommand{\nomgroup}[1]{%
 \ifthenelse{\equal{#1}{O}}{\item[\textit{Operators}]}{%
 \ifthenelse{\equal{#1}{I}}{\item[\textit{Indices}]}{%
 \ifthenelse{\equal{#1}{A}}{\item[\textit{Acronyms}]}{%
 `\ifthenelse{\equal{#1}{V}}{\item[\textit{Variables and parameters}]}{}}}}}
\usepackage{scalerel}
\usetikzlibrary{svg.path}
\definecolor{orcidlogocol}{HTML}{A6CE39}
\tikzset{
 orcidlogo/.pic={
 \fill[orcidlogocol] svg{M256,128c0,70.7-57.3,128-128,128C57.3,256,0,198.7,0,128C0,57.3,57.3,0,128,0C198.7,0,256,57.3,256,128z};
 \fill[white] svg{M86.3,186.2H70.9V79.1h15.4v48.4V186.2z}
 svg{M108.9,79.1h41.6c39.6,0,57,28.3,57,53.6c0,27.5-21.5,53.6-56.8,53.6h-41.8V79.1z M124.3,172.4h24.5c34.9,0,42.9-26.5,42.9-39.7c0-21.5-13.7-39.7-43.7-39.7h-23.7V172.4z}
 svg{M88.7,56.8c0,5.5-4.5,10.1-10.1,10.1c-5.6,0-10.1-4.6-10.1-10.1c0-5.6,4.5-10.1,10.1-10.1C84.2,46.7,88.7,51.3,88.7,56.8z};
 }
}

\newcommand\orcidicon[1]{\href{https://orcid.org/#1}{\mbox{\scalerel*{ \begin{tikzpicture}[yscale=-1,transform shape]
 \pic{orcidlogo};
 \end{tikzpicture}
 }{|}}}}

\usepackage{multirow}
\usepackage{soul}
\usepackage{algorithm}
\usepackage{algpseudocode}
\usepackage{graphicx}
\usepackage{dsfont}
\usepackage{xcolor}
\usepackage{cite}

\usepackage{graphicx}
\usepackage{changepage}
\usepackage{subcaption}
\usepackage{placeins}
\usepackage[utf8]{inputenc}
\usepackage{mathtools}
\usepackage[mathscr]{euscript}
\usepackage{amsmath}
\usepackage{amssymb}
\usepackage{amsfonts}
\usepackage{amssymb}
\usepackage{amsmath}
\usepackage{verbatim}
\usepackage[T1]{fontenc}
\usepackage[utf8]{inputenc}
\usepackage{latexsym}
\usepackage{enumerate}
\usepackage{indentfirst}
\usepackage{calligra}
\usepackage{ulem}
\usepackage{graphicx}
\usepackage{array}
\usepackage{float}
\usepackage{bbm}

\usepackage{mleftright}
\usepackage{colortbl}
\makeatletter
\let\old@ps@headings\ps@headings
\let\old@ps@IEEEtitlepagestyle\ps@IEEEtitlepagestyle
\def\psccfooter#1{%
 \def\ps@headings{%
 \old@ps@headings%
 \def\@oddfoot{\strut\hfill#1\hfill\strut}%
 \def\@evenfoot{\strut\hfill#1\hfill\strut}%
 }%
 \def\ps@IEEEtitlepagestyle{%
 \old@ps@IEEEtitlepagestyle%
 \def\@oddfoot{\strut\hfill#1\hfill\strut}%
 \def\@evenfoot{\strut\hfill#1\hfill\strut}%
 }%
 \ps@headings%
}
\makeatother

\usepackage{soul}
\usepackage{multirow}
\usepackage{soul,color}
\usepackage{graphicx}
\usepackage{dsfont}
\usepackage{subcaption}

\usepackage[utf8]{inputenc}
\usepackage{mathtools}
\usepackage[mathscr]{euscript}
\usepackage{amsmath}
\usepackage{amssymb}
\usepackage{amsfonts}
\usepackage{verbatim}
\usepackage[T1]{fontenc}
\usepackage[utf8]{inputenc}

\usepackage{latexsym}
\usepackage{enumerate}
\usepackage{indentfirst}
\usepackage{calligra}
\usepackage{ulem}
\usepackage{multirow}

\usepackage{array}
\usepackage{float}
\usepackage{bbm}

\usepackage{eurosym}

\usepackage{hyperref}

\usepackage{caption}
\usepackage{tikz}
\usepackage{pgfplots}

\pgfplotsset{compat=1.8}
\usepgfplotslibrary{statistics}
\pgfmathdeclarefunction{fpumod}{2}{%
 \pgfmathfloatdivide{#1}{#2}%
 \pgfmathfloatint{\pgfmathresult}%
 \pgfmathfloatmultiply{\pgfmathresult}{#2}%
 \pgfmathfloatsubtract{#1}{\pgfmathresult}%
 \pgfmathfloatifapproxequalrel{\pgfmathresult}{#2}{\def\pgfmathresult{5}}{}%
 }

\usepackage{tabularx}
\usepackage{flushend}
\usepackage{textcomp}
\usepackage{xcolor}
\usepackage{import}

\usepackage{csvsimple}
\usepackage{tabularx}
\usepackage{adjustbox}

\usetikzlibrary{trees,decorations,shadows}
\tikzset{level 1/.style={sibling angle=45,level distance=4mm}}
\usetikzlibrary{arrows.meta}
\usepackage{forest}
\usetikzlibrary{external}

\let\oldtikzexternalgetnextfilename\tikzexternalgetnextfilename \renewcommand{\tikzexternalgetnextfilename}[1]{\oldtikzexternalgetnextfilename{#1}\expandafter\tikzsetnextfilename\expandafter{#1}}

\usepackage{pgfplotstable}
\usepackage[outline]{contour}
\contourlength{1.2pt}

\pgfplotsset{compat=1.13} 

\usetikzlibrary{spy}
\usetikzlibrary{calc}
\usetikzlibrary{fadings}
\usetikzlibrary{patterns}
\usetikzlibrary{shadows}
\usetikzlibrary{mindmap}
\usetikzlibrary{backgrounds}
\usetikzlibrary{shapes.symbols}
\usetikzlibrary{shapes.multipart}
\usetikzlibrary{shapes.geometric}
\usetikzlibrary{automata,positioning}
\usetikzlibrary{decorations.fractals} 
\usetikzlibrary{decorations.markings}
\usetikzlibrary{decorations.pathreplacing}
\usetikzlibrary{decorations.pathmorphing}
\usepackage{svg}
 
\usepackage{bm}

\usepackage{nomencl}
\makenomenclature
\tikzset{edge from parent/.style={segment angle=10,draw}}

\tikzset{
 my rounded corners/.append style={rounded corners=2pt},
}

\def\BibTeX{{\rm B\kern-.05em{\sc i\kern-.025em b}\kern-.08em
 T\kern-.1667em\lower.7ex\hbox{E}\kern-.125emX}}

\renewcommand{\nomgroup}[1]{%
 \ifthenelse{\equal{#1}{O}}{\item[\textit{Operators}]}{%
 \ifthenelse{\equal{#1}{I}}{\item[\textit{Indices}]}{%
 \ifthenelse{\equal{#1}{A}}{\item[\textit{Acronyms}]}{%
 `\ifthenelse{\equal{#1}{V}}{\item[\textit{Variables and parameters}]}{}}}}}
\usepackage{scalerel}
\usetikzlibrary{svg.path}
\definecolor{orcidlogocol}{HTML}{A6CE39}
\tikzset{
 orcidlogo/.pic={
 \fill[orcidlogocol] svg{M256,128c0,70.7-57.3,128-128,128C57.3,256,0,198.7,0,128C0,57.3,57.3,0,128,0C198.7,0,256,57.3,256,128z};
 \fill[white] svg{M86.3,186.2H70.9V79.1h15.4v48.4V186.2z}
 svg{M108.9,79.1h41.6c39.6,0,57,28.3,57,53.6c0,27.5-21.5,53.6-56.8,53.6h-41.8V79.1z M124.3,172.4h24.5c34.9,0,42.9-26.5,42.9-39.7c0-21.5-13.7-39.7-43.7-39.7h-23.7V172.4z}
 svg{M88.7,56.8c0,5.5-4.5,10.1-10.1,10.1c-5.6,0-10.1-4.6-10.1-10.1c0-5.6,4.5-10.1,10.1-10.1C84.2,46.7,88.7,51.3,88.7,56.8z};
 }
}

\usepackage{mdframed}

\newmdenv[leftline=false,rightline=false,linewidth=1pt]{topbot2}

\begin{document}

\title{Consumer empowerment: need, definition,  quantification, outcomes and challenges}
\author{Md Umar Hashmi,\\
KU Leuven \& EnergyVille, Belgium}
\date{}

\maketitle

\begin{abstract}
The transition toward decentralized, digitalized, and low-carbon electricity systems is fundamentally changing the role of electricity consumers. Traditionally passive end-users are increasingly expected to integrate distributed energy resources (DERs) and participate in energy management, local energy markets, and demand response, thus, facilitating active grid support. This paper examines consumer empowerment (CE) in electricity networks with a specific focus on small consumers connected to distribution networks. The paper first discusses the technical, economic, and societal drivers underlying the growing need for CE, including increasing DER penetration, electrification of heating and transportation, network congestion, and flexibility requirements. A comprehensive definition of CE is then proposed based on six elements: active participation, enhanced choice, information and control, automation and controllability, self-sufficiency and resilience, and energy communities with shared ownership. 
The strength of the proposed CE definition is an explicit consideration of weak-grid (isolated power networks, microgrids, etc.) and Global South power networks where security of supply is not guaranteed.
To support consistent assessment across different electricity systems, a framework for quantifying CE is developed based on the availability of technology and policy enablers and consumers' ability to ensure security of supply in the absence of a reliable grid.
The paper further reviews international experiences and implementation practices across different continents, highlighting variations in technological readiness, regulatory support, and security-of-supply considerations. Finally, the expected outcomes and challenges of CE are 
discussed. 
The expected outcomes of CE are
improving
affordability, flexibility, resilience, market efficiency, DER integration, and decarbonization, while its successful implementation requires overcoming barriers related to information access, digital literacy, affordability, regulation, interoperability, privacy, cybersecurity, and consumer engagement. CE is a key enabler of future consumer-centric electricity systems and helps manage the energy transition economically by avoiding massive infrastructural upgrades that would otherwise be needed.

\end{abstract}

\vspace{18pt}

\begin{IEEEkeywords}
Distribution networks, Consumer empowerment, electricity consumers, affordability, power grid.
\end{IEEEkeywords}


\pagebreak

\setcounter{tocdepth}{2}
\tableofcontents

\pagebreak

\section*{Abbreviations}

\begin{align*}
\text{ACER} & \quad \text{Agency for the Cooperation of Energy Regulators} \\
\text{AMI} & \quad \text{Advanced metering infrastructure} \\
\text{BESS} & \quad \text{Battery energy storage systems} \\
\text{CE} & \quad \text{Consumer empowerment} \\
\text{DER} & \quad \text{Distributed energy resources} \\
\text{DG} & \quad \text{Distributed generation} \\
\text{DN} & \quad \text{Distribution network} \\
\text{DR} &  \quad \text{Demand response} \\
\text{DSO} & \quad \text{Distribution system operator} \\
\text{EMD} & \quad \text{Energy Market Design} \\
\text{EU} & \quad \text{European Union} \\
\text{EV} & \quad \text{Electric Vehicle} \\
\text{FERC} & \quad \text{Federal Energy Regulatory Commission} \\
\text{HEMS} & \quad \text{Home energy management system} \\
\text{HP} & \quad \text{Heat pump} \\
\text{IEA} & \quad \text{International Energy Agency} \\
\text{IoT} & \quad \text{Internet of Things} \\
\text{P2P} & \quad \text{Peer-to-peer} \\
\text{PV} & \quad \text{Photovoltaic} \\
\text{RES} & \quad \text{Renewable energy sources} \\
\text{SME} & \quad \text{Small-to-medium-sized enterprise} \\
\text{ToU} & \quad \text{Time-of-use tariff} \\
\text{VPP} & \quad \text{Virtual power plant} \\
%
%
%
%
%
\end{align*}

\pagebreak

\section{Introduction: Need for Consumer Empowerment}
\label{section1}

\begin{figure}[htbp!]
    \centering
    \includegraphics[width=0.768\textwidth]{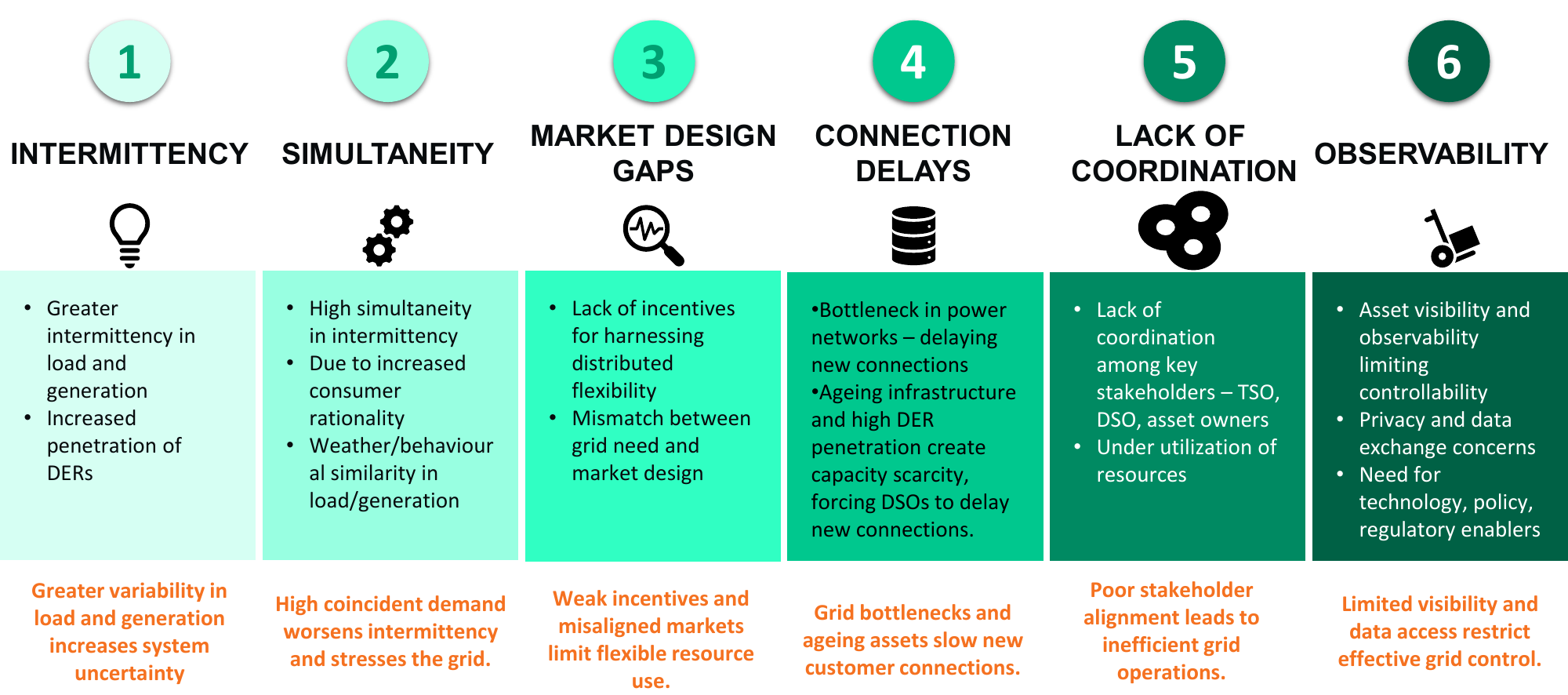} 
    \vspace{-10pt}
    \caption{\small{Some of the key challenges in distribution networks.}}
    \label{fig:challenges}
\end{figure}

The challenges emerging in modern distribution networks (DNs) are closely linked with an increasing penetration of distributed energy resources (DERs), electric vehicles (EVs), and electrified heating, which contribute to intermittency, simultaneity of demand, network congestion, and reduced observability; traditional network reinforcement alone becomes increasingly costly and insufficient (see Fig. \ref{fig:challenges}). In particular, the variability of distributed generation (DG), coincident consumer behaviour, market design gaps, connection bottlenecks, and limited coordination among stakeholders all highlight the growing importance of leveraging flexibility at the consumer level. Empowered consumers, equipped with access to information, smart technologies, dynamic tariffs, and market participation mechanisms, can actively adjust consumption patterns, provide demand-side flexibility, coordinate distributed resources, and support local grid operation. By transforming small consumers from passive electricity users into active participants in the energy system, consumer empowerment (CE) can help mitigate peak demand, reduce network congestion, improve the integration of renewable energy sources (RES), enhance system observability through data sharing, and ultimately increase the efficiency and resilience of DNs.

\vspace{-15pt}
Historically, electricity consumers were passive recipients of electricity supplied by centralized generation resources. However, the ongoing transition toward decentralized, digitalized, and low-carbon energy systems requires consumers to become active participants in energy management, flexibility provision, and RES integration. The rapid growth of DERs and electrification of transportation and heating have shifted a significant portion of decision-making and operational capabilities toward end-users. Consequently, consumers must be provided with the information, technologies, market access, and control mechanisms necessary to effectively interact with increasingly complex electricity systems \cite{iea_customer_choice,doe_der}. 
Furthermore, Geelen et al. in \cite{geelen2013empowering} recommend that CE should extend beyond technological solutions and financial incentives by improving user interaction with smart energy systems, supporting behavioural change through targeted interventions, and enabling community-based energy management and participation. 
\begin{figure}[htbp!]
    \centering
    \includegraphics[width=0.7\textwidth]{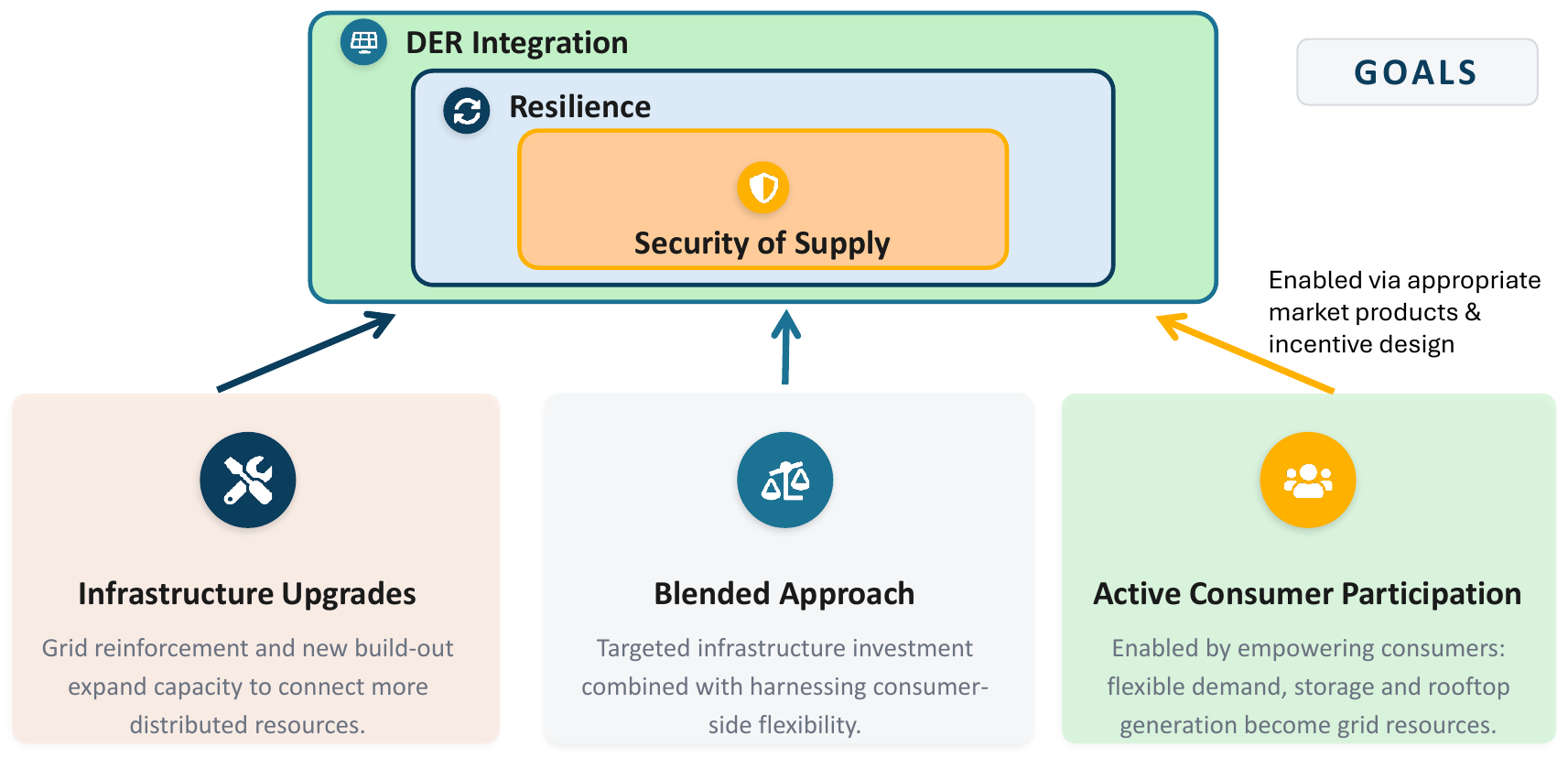} 
    \vspace{-5pt}
    \caption{\small{Pathways to the Power Network Transition.}}
    \label{fig:approached}
\end{figure}
Figure~\ref{fig:approached} illustrates how the ongoing transition of electricity networks, driven by increasingly ambitious objectives ranging from (a) security of supply to (b) resilience and (c) large-scale DER integration, requires a corresponding transformation in the role of consumers. While these goals can be achieved through conventional infrastructure upgrades,
an economically viable alternative is 
active consumer participation, supported by appropriate market mechanisms and incentives that enable consumers to provide flexibility for grid services while increasing energy affordability for themselves. 
The third approach is a blended solution in which infrastructural upgrades and consumer participation are both used for ensuring operation goals of the power network are met.

\textit{Electricity consumer} 
differ significantly in size, technical capabilities, and network integration. Large industrial consumers are frequently connected directly to the transmission network and often perform functions comparable to those of distribution network operators within their premises, including the management of internal distribution infrastructure and energy resources \cite{hashmi2026network, Borenstein2016, Kirschen2018}. A second category comprises commercial and small-to-medium-sized enterprises (SMEs), which are typically connected at the medium-voltage level and possess greater operational flexibility and contractual sophistication than residential users \cite{JamasbPollitt2008}. For both industrial and commercial customers, the technical, operational, and contractual aspects of network access are generally well established through grid codes, connection agreements, and bilateral contractual arrangements. Consequently, \textbf{this work focuses on small electricity consumers connected to DNs}, including residential customers and small businesses. 
These consumers represent the largest and most diverse customer segment and are increasingly expected to participate actively in the energy transition through DERs, demand response (DR), and flexibility services \cite{ParagSovacool2016,Burger2019}. However, they often face greater informational, technical, and economic barriers to effective participation in electricity markets and flexibility mechanisms \cite{RosenowEyre2016,StadelmannSteffen2018}. Their growing role in DNs makes them particularly relevant for this study on CE and active network participation.

CE is crucial for \textbf{maintaining power system reliability} and flexibility for accommodating increased intermittency and simultaneity, as shown in DN challenges in Fig. \ref{fig:challenges}.
As variable RES constitute an increasing share of electricity generation, power systems require additional flexibility to balance supply and demand. Active consumer participation through DR, flexible consumption, distributed battery energy storage systems (BESS), and automated energy management can provide cost-effective flexibility services while reducing the need for expensive network reinforcements and backup generation assets (see Fig.~\ref{fig:approached}). The International Energy Agency (IEA) highlights that customer participation and DR are critical enablers of power system flexibility, electricity security, and market efficiency \cite{iea_customer_choice,iea_electricity2026}.

Another major driver for CE is the growing \textbf{need to improve affordability} and reduce consumer energy costs. Across many regions, consumers face rising electricity demand, increasing electrification of transport and heating, and growing investment requirements for grid modernization. Empowered consumers can respond to dynamic tariffs, optimize self-consumption from DG, participate in virtual power plants (VPPs), and receive compensation for flexibility services. In the European Union, recent Energy Market Design (EMD) reforms explicitly promote consumer participation, energy sharing, and greater supplier choice to improve affordability and protect consumers from price volatility \cite{europaProtectingEmpowering,ec_market_design}. Similarly, studies in the United States show that DERs and VPPs can reduce system costs while lowering consumer electricity bills \cite{pew_der,doe_der}. 
For achieving stable and affordable energy prices, the European Commission recommends (i) a variety of electricity contracts, (ii) better information on energy offers, (iii) ensuring continuity of supply and (iv) access to hedging products facilitating self-consumption within an energy community \cite{europaProtectingEmpowering}.

CE is further needed to \textbf{accelerate decarbonization} and the global energy transition. Electrification, energy efficiency, digitalization, and decentralization all depend heavily on consumer decisions. Consumers determine whether to adopt energy-efficient appliances, rooftop photovoltaic (PV) systems, heat pumps (HPs), and BESS technologies. Consumers International emphasizes that demand-side actions could contribute substantially to global emissions reductions, but consumers face technological, financial, informational, and institutional barriers that must be addressed through empowerment measures \cite{ci_clean_energy}. The organization further argues that consumers should not be viewed as 
active agents capable of driving transformational change when provided with appropriate support and protections \cite{ci_empowerment,enel_ci}.

Finally, CE \textbf{enhances energy resilience and community participation}. 
Recent experiences with extreme weather events, natural disasters, and energy price crises have highlighted vulnerabilities associated with highly centralized electricity systems. Increasing frequency and intensity of climate-related events such as heatwaves, cold waves, wildfires, cyclones, and floods have emerged as major causes of large-scale power outages worldwide, emphasizing the need for more resilient and decentralized energy systems \cite{iea_climate_resilience,iea_power_transition,ipcc_extremes, panteli2026fundamentals}. In parallel, the global energy crisis following Russia's invasion of Ukraine demonstrated how dependence on centralized fuel supply chains can expose consumers to severe electricity price volatility and affordability challenges, further motivating interest in DERs, local energy communities, and consumer self-sufficiency \cite{iea_electricity_market_2022,iea_climate_security}.
Empowered consumers equipped with local generation, BESS, microgrids, and energy management systems are better positioned to maintain critical energy services during disruptions. In addition, energy communities, peer-to-peer (P2P) trading, and shared ownership models enable consumers to collectively enhance local energy security and create additional social and economic value \cite{europaProtectingEmpowering,pew_der}. 

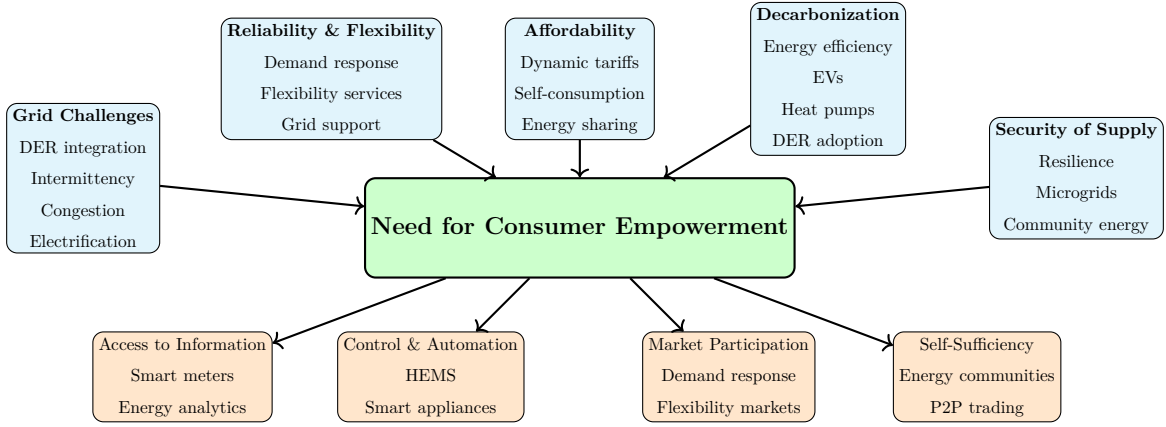
\begin{figure*}[!t] \centering \begin{tikzpicture}[ box/.style={ draw, rounded corners, align=center, minimum width=3cm, minimum height=1.2cm, fill=cyan!10 }, arrow/.style={ ->, thick }, scale=0.657, transform shape ] 
\node[box] (grid) at (0,2.5) { \textbf{Grid Challenges}\\ DER integration\\ Intermittency\\ Congestion\\ Electrification }; \node[box] (reliability) at (5,4.5) { \textbf{Reliability \& Flexibility}\\ Demand response\\ Flexibility services\\ Grid support }; \node[box] (affordability) at (10,4.5) { \textbf{Affordability}\\ Dynamic tariffs\\ Self-consumption\\ Energy sharing }; \node[box] (decarbonization) at (15,4.5) { \textbf{Decarbonization}\\ Energy efficiency\\ EVs\\ Heat pumps\\ DER adoption }; \node[box] (resilience) at (20,2.5) { \textbf{Security of Supply}\\ Resilience\\ Microgrids\\ Community energy }; 
\node[draw, thick, fill=green!20, rounded corners, minimum width=7cm, minimum height=2cm] (emp) at (10,1.5) { \Large\textbf{Need for Consumer Empowerment} }; 
\node[box, fill=orange!20] (info) at (2,-1.5) { Access to Information\\ Smart meters\\ Energy analytics }; \node[box, fill=orange!20] (control) at (7,-1.5) { Control \& Automation\\ HEMS\\ Smart appliances }; \node[box, fill=orange!20] (market) at (13,-1.5) { Market Participation\\ Demand response\\ Flexibility markets }; \node[box, fill=orange!20] (community) at (18,-1.5) { Self-Sufficiency\\ Energy communities\\ P2P trading }; 
\draw[arrow] (grid) -- (emp); \draw[arrow] (reliability) -- (emp); \draw[arrow] (affordability) -- (emp); \draw[arrow] (decarbonization) -- (emp); \draw[arrow] (resilience) -- (emp); 
\draw[arrow] (emp) -- (info); \draw[arrow] (emp) -- (control); \draw[arrow] (emp) -- (market); \draw[arrow] (emp) -- (community); \end{tikzpicture} \caption{\small{Drivers for the growing need for consumer empowerment and the resulting consumer capabilities.} } \label{fig:consumer_empowerment_need} \end{figure*}

Figure~\ref{fig:consumer_empowerment_need} illustrates the need for adopting CE and the implications of such adoption on the capabilities consumers will have. This illustration summarizes the section on "\textit{Why CE is needed?}".


\subsection{Scope of the Paper}
Consumer empowerment has emerged as a key enabler of modern electricity networks, driven by the increasing deployment of DERs, digital technologies, and consumer-centric market designs. While the concept is widely recognized in policy, regulatory, and academic discussions, its interpretation, measurement, and implications for power system operation remain fragmented. Furthermore, CE encompasses a broad range of dimensions, including access to information, choice, control, active participation, automation, resilience, and community engagement, making it challenging to establish a common understanding across stakeholders. The focus of this work is towards CE for small electricity consumers, i.e., residential or commercial consumers typically connected to the low-voltage electricity networks.

This paper addresses these challenges by providing a comprehensive examination of CE in the context of power networks. Specifically, the paper investigates the \textbf{need for CE} and the drivers behind its growing importance in the energy transition. It proposes a new \textbf{definition} of CE based on the literature and its key attributes.
The paper further explores approaches for \textbf{quantifying CE}, and proposes new metrics for its quantification.
Finally, the key \textbf{challenges} are identified and \textbf{expected outcomes} of implementing/facilitating CE in power networks are discussed.

\textit{Who benefits from CE?} 
Consumers can benefit through lower electricity bills, improved resilience, greater choice, and access to flexibility markets, while DSOs, TSOs, market operators, regulators, and policy makers may benefit from improved system efficiency, reduced network investments, enhanced market competition, and progress toward decarbonization objectives. However, benefits are not always distributed equally; consumers lacking access to DERs, financing, or digital capabilities may be unable to participate, raising equity concerns and potentially creating unintended distributional effects \cite{Sorrell2007}. Therefore, CE policies should be designed to balance efficiency, affordability, and fairness across all stakeholder groups.
The scope of this paper does not delve into fairness within consumers with and without DERs.

\begin{tcolorbox}[colback=gray!8!white,colframe=blue!75!black]
\small{
Maximization of consumer empowement can be considered as a \textbf{societal goal} as it leads to improved utilization of power networks while making energy affordable for consumers.
}
\end{tcolorbox}

\begin{tcolorbox}[colback=red!8!white,colframe=red!75!red]
\small{
This paper attempts to answer the following questions:
\begin{itemize}
    \item What are the needs for CE?
    \item What is the definition of CE?
    \item How to quantify CE? A quantification that is not just applicable to reliable first-world power networks but also to the less reliable Global South countries.
    \item What are the challenges associated with CE?
    \item What are the expected outcomes of enabling/facilitating CE?
    \item What are the use cases and real-world experiences of CE implementations?
\end{itemize}
}
\end{tcolorbox}

\subsection{Contributions} 
\begin{tcolorbox}[colback=green!8!white,colframe=green!75!green]
\small{
The main contributions of this paper are summarized as follows: 
\begin{itemize} 
\item \textbf{A pragmatic definition of consumer empowerment} (CE) is proposed that is applicable to both advanced electricity systems and regions with weak or unreliable grids, thereby extending the concept beyond its traditional focus on developed economies. 
\item A \textbf{framework for quantifying CE} is introduced based on three key enabling dimensions: (i) availability of technological enablers, (ii) access to policy and regulatory enablers, and (iii) the ability to ensure security of supply. 
\item The paper systematically identifies and discusses the major \textbf{challenges and expected outcomes of CE}, covering technical, economic, regulatory, behavioural, societal, and power system perspectives. 
\item A \textbf{collection of real-world examples} from different regions of the world is presented, illustrating how CE is realized through technology deployment, supportive policies, energy communities, and resilience-oriented initiatives. 
\end{itemize}

}
\end{tcolorbox}

This paper is organized as follows. Section~\ref{section1} provides an introduction to the paper and discusses the need for CE in modern electricity networks and defines the scope and contributions of the paper. Section~\ref{section2} presents a pragmatic definition of CE and identifies its key dimensions. Section~\ref{section3} introduces a framework for quantifying CE and proposes relevant metrics for its assessment. Section~\ref{section4} reviews the state of CE worldwide and presents real-world implementation examples. Section~\ref{section5} discusses the expected outcomes and benefits of CE, while Section~\ref{section6} examines the key challenges and barriers to its realization. Finally, Section~\ref{section7} concludes the paper and outlines directions for future research.

\pagebreak

\pagebreak

\section{Consumer empowerment: proposed definition}
\label{section2}

\begin{figure}[htbp!]
    \centering
    \includegraphics[width=0.603\textwidth]{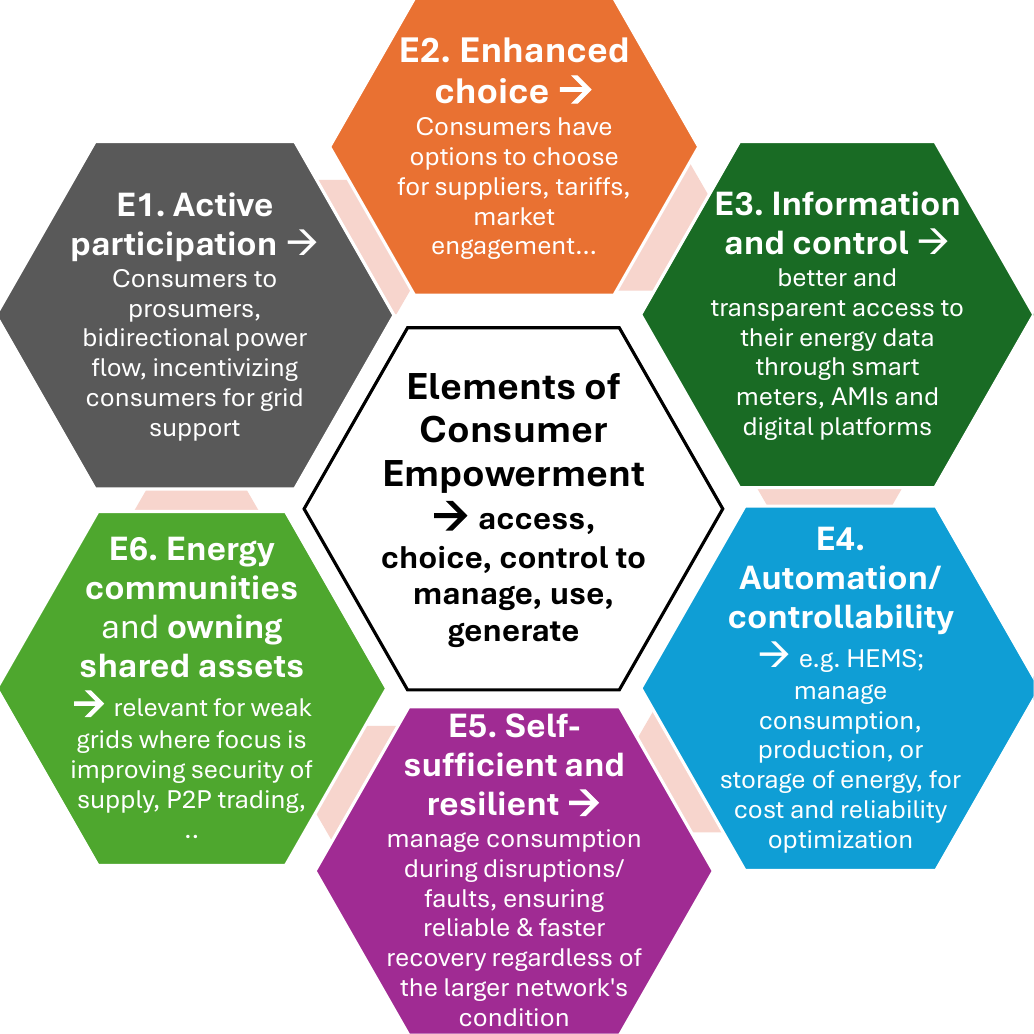} 
    \vspace{-5pt}
    \caption{\small{Elements of consumer empowerment.}}
    \label{fig:whatCE}
\end{figure}

Consumer empowerment in electricity networks marks a paradigm shift from passive consumption to active participation, granting individuals the access, choice, and control to manage, use, and generate their own energy. This transformation is built on several interconnected elements: consumers evolve into "\textit{prosumers}" with bidirectional power flows, gain enhanced choice over suppliers and tariffs, and leverage information from smart meters alongside automation tools like Home Energy Management Systems (HEMS) to optimize costs and reliability \cite{whittle2020empowering}. For weaker grids, this empowerment takes on added significance through energy communities that own shared assets, enabling P2P trading and bolstering security of supply, while self-sufficiency and resilience allow consumers to adjust consumption during disruptions—ensuring reliable power and faster recovery regardless of the larger network's condition. Together, these elements redefine the consumer's role as an active, resilient partner in the energy system.
Fig.\ref{fig:whatCE} visualizes the six elements (E1 to E6) of CE and is detailed next.
Elements E1 to E6 denote the attributes of CE.
The presence and extent of these attributes may vary across different power networks.

\subsection{E1: Facilitating active consumer participation}

Active participation transforms consumers into prosumers who can generate, store, and export their own electricity, enabling bidirectional power flows where energy moves both from the grid to the user and back from the user to the grid. This shift gives consumers a direct role in grid operations, allowing them to not only meet their own needs but also supply surplus energy and support local grid stability \cite{shaukat2018survey}.

Empowered consumers can adjust their consumption in response to price or reliability signals, participate in DR programs, and, where technically enabled, aggregate with other consumers to gain market influence. DR programmes are reported to enhance engagement by incentivising shifts in electricity use in reaction to pricing or system needs, fostering both economic benefit and active involvement in energy management \cite{shaukat2018survey}. Analytical work on DR contracts further stresses that engaging and empowering small, distribution-connected consumers requires policy design attentive to consumer preferences, since many consumers favour simplicity over complex dynamic programmes \cite{he2013engage}.

\textit{How does no E1 in CE looks like?}: In many power networks, consumers receive a fixed electricity billing framework, and there are no incentives for consumers to actively participate in ensuring operation integrity of DNs \cite{colombia_tariff_distortions_2021}. E1 attribute depends also on E3 and E4, as if there is no monitoring capability, active participation cannot be realized.

\subsection{E2: Consumer enhanced choices}
In DNs, enhanced choice empowers consumers to act as active market participants by selecting from multiple suppliers, dynamic pricing plans, and green energy options, while also deciding whether to generate, store, or sell their own power. This transforms the consumer from a passive electricity consumer into a decision-maker who can align their energy choices with their financial goals, environmental values, and lifestyle needs.

\begin{figure}[htbp!]
    \centering
    \includegraphics[width=0.83\textwidth]{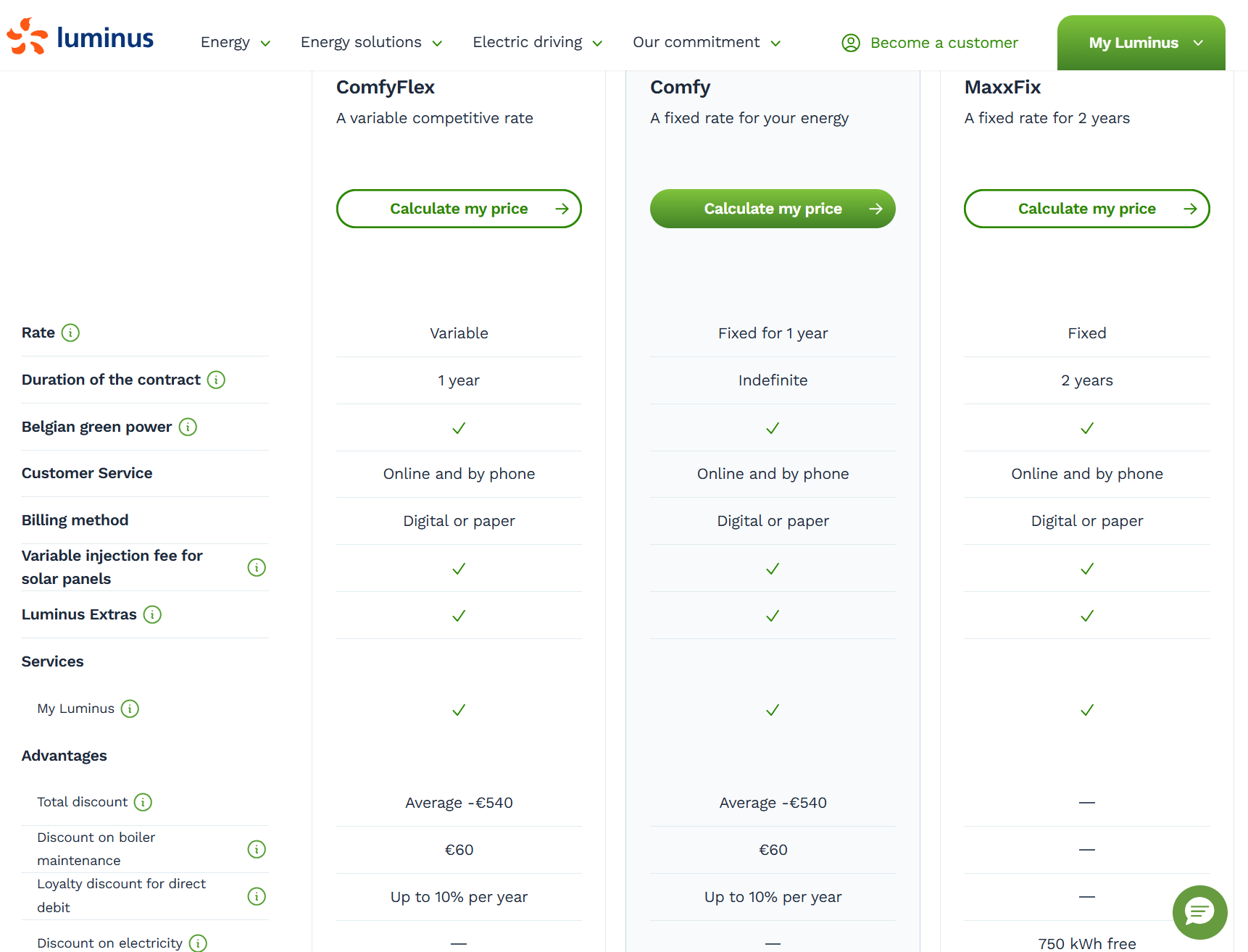} 
    \vspace{-5pt}
    \caption{\small{Consumer contracts to choose from by Luminous in Flanders, Belgium \cite{luminusOnzeBeste}.}}
    \label{fig:E2_choice}
\end{figure}

Figure~\ref{fig:E2_choice} illustrates the range of electricity supply contracts offered by Luminus to residential consumers in Flanders, Belgium. The availability of multiple tariff options, including fixed-price and variable-price contracts with different contract durations and service features, demonstrates how modern retail electricity markets can provide consumers with greater choice, enabling them to select energy products that best match their preferences, risk tolerance, and consumption patterns \cite{luminusOnzeBeste}.

\textit{How does no E2 in CE looks like?}: Consumers do not have choice/options for selecting contracts (tariffs), suppliers, etc.

\subsection{E3: Access and control of consumer energy analytics}
A foundational element of CE is access to accurate, timely information about consumption and costs. Advanced Metering Infrastructure (AMI) and real-time monitoring allow consumers to observe and manage their own energy profiles, which is considered a primary enabler of CE in smart grid systems \cite{shaukat2018survey}. 
Information and control empower consumers by granting them real-time access to their granular energy usage data via smart meters and digital platforms, enabling them to actively monitor, manage, and optimize their consumption patterns. This transparency transforms passive billing into proactive energy management, giving consumers the ability to make informed decisions that reduce costs, shift usage to off-peak times, or respond to grid conditions.
Policy proposals in this area typically include minimum requirements for billing information, wider deployment of smart meters, and more accurate measurement of actual consumption \cite{ConsumerPolicy2018}.

\begin{figure}[htbp!]
    \centering
    \includegraphics[width=0.8\textwidth]{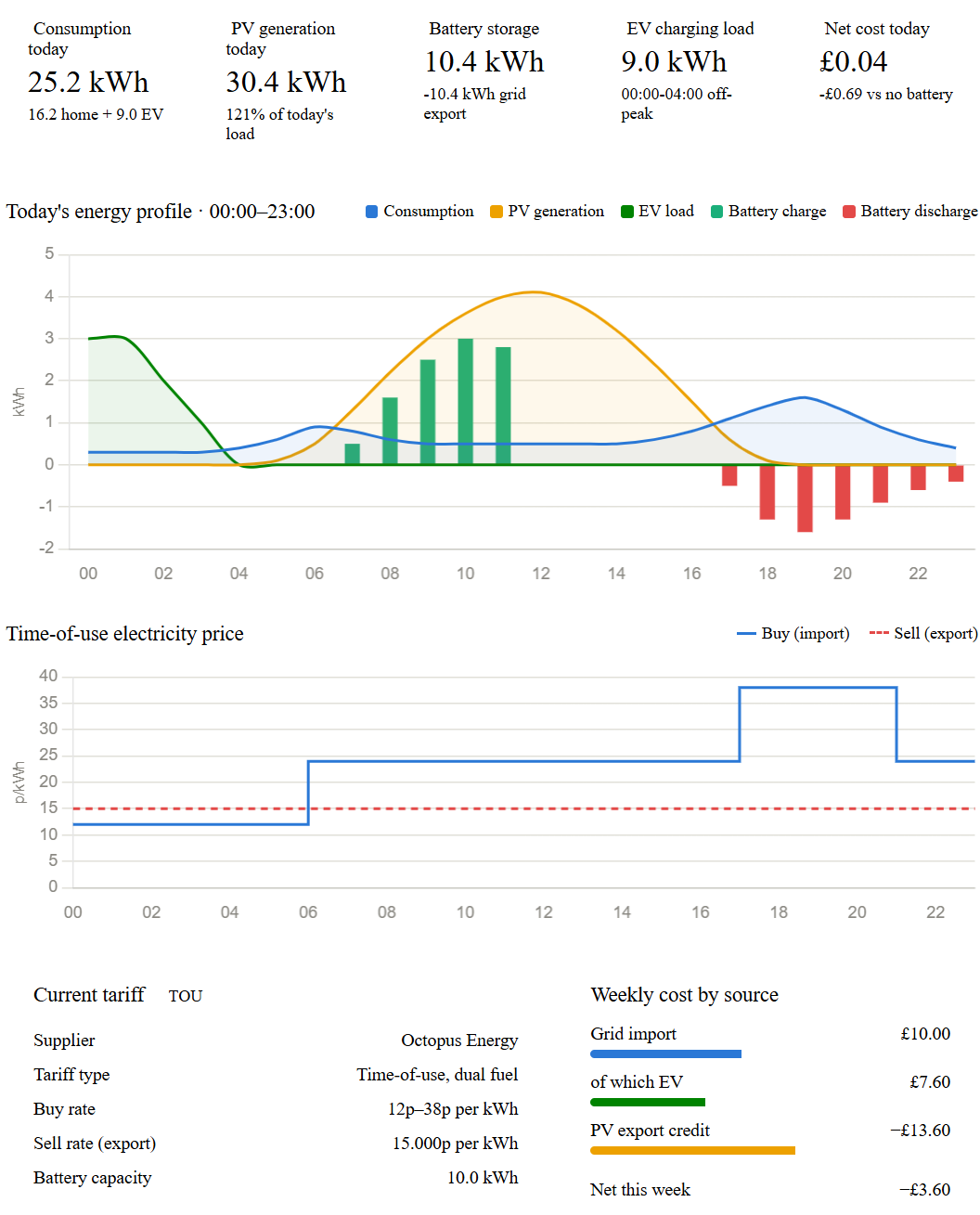} 
    \vspace{-5pt}
    \caption{\small{A stylized consumer energy dashboard providing visibility and control of household energy assets, including its consumption, PV generation, battery storage, EV charging, electricity prices, and energy costs.}}
    \label{fig:E3_dashboard}
\end{figure}

Figure~\ref{fig:E3_dashboard} illustrates a stylized consumer energy dashboard that provides visibility and control of DERs and household loads. The dashboard presents key daily indicators, including electricity consumption, PV generation, battery operation, EV charging demand, and electricity costs. In addition, consumers can monitor energy flows, local asset performance, and time-of-use (ToU) electricity prices, enabling informed operational decisions. Such dashboards are an important technological enabler of CE, allowing consumers to optimize energy use, increase self-consumption, reduce costs, and actively participate in modern DNs.

\textit{How does no E3 in CE looks like?}: Consumers do not have access to their consumption/generation data.

\subsection{E4: Automation and Controllability}
Automation and controllability empower consumers by using technologies such as HEMS to autonomously schedule and dispatch DERs, including PV systems, BESS, HPs, and EVs, in response to electricity prices, user preferences, and grid conditions \cite{HEMSReview2019,HEMSOptimization2021}. This allows consumers to optimize electricity costs, increase self-consumption of locally generated renewable energy, reduce peak demand, and provide flexibility services to the electricity network with minimal user intervention. Furthermore, automated control reduces the complexity associated with managing multiple energy assets and enables consumers to participate in demand response and flexibility programs without requiring continuous manual decision-making \cite{shaukat2018survey,he2013engage}.

Modern HEMS leverage smart meter data, forecasts of demand and renewable generation, dynamic electricity tariffs, and user-defined preferences to determine optimal operating schedules for controllable loads and distributed resources \cite{HEMSReview2019,HEMSOptimization2021}. Beyond economic benefits, HEMS can improve consumer comfort, increase system reliability, facilitate the integration of higher levels of DERs, and support network operation through the provision of flexibility and ancillary services. As electricity systems become increasingly digitalized and decentralized, HEMS are expected to play a central role in enabling automated and scalable consumer participation in future smart grids \cite{ParagSovacool2016,Burger2019}.


\textit{How does no E4 in CE looks like?}: In the absence of E4, consumers have limited ability to automate and optimize the operation of DERs and flexible loads. In many regions, DER adoption remains low or insufficiently incentivized, while HEMS and other automation technologies are not yet widely deployed, requiring consumers to manually manage energy consumption and reducing the potential for flexibility provision and active participation in electricity markets.

\subsection{E5: Self-sufficient and resilient electricity consumption}
CE, in terms of self-sufficiency and resilience, means shifting consumers from passive users to active "\textit{prosumers}" who generate, store, and manage their own energy to reduce dependency on the central grid, while also having the ability to island themselves or autonomously adjust their consumption during disruptions, ensuring reliable power and faster recovery regardless of the larger network's condition \cite{ParagSovacool2016,Morstyn2018}.
This dimension of empowerment is particularly important in electricity systems with high penetrations of distributed energy resources, where local generation and storage can support both consumer needs and system operation. Through technologies such as rooftop PV systems, BESS, and HEMS, consumers can increase self-consumption, reduce exposure to electricity price volatility, and improve their ability to maintain critical loads during outages \cite{Mengelkamp2018}. Furthermore, local energy communities, P2P energy trading platforms, and community microgrids enable groups of consumers to collectively improve security of supply and energy autonomy while reducing reliance on centralized infrastructure \cite{Mengelkamp2018}. Consequently, self-sufficiency and resilience extend CE beyond economic optimization by enhancing the capability of consumers to maintain reliable electricity supply under both normal and emergency operating conditions.

This attribute of CE is particularly important in regions with weak or unreliable electricity infrastructure, where consumers experience frequent outages, load shedding, or limited grid availability \cite{jana_power_2021}. In many countries of the Global South, as well as in isolated microgrids and remote communities, self-sufficiency and resilience can be more valuable than market participation, as they directly improve security of supply and enable consumers to maintain critical energy services during grid disruptions.


\textit{How does no E5 in CE looks like?}: Consumers have no alternative to ensure security of supply.

\subsection{E6: Energy communities and shared assets}

In weak grids, CE through the formation of energy communities, households collectively own and operate shared assets such as solar mini-grids, BESS, or community energy resources \cite{monergirona2022microgrids}, enabling them to reduce dependence on unreliable centralized infrastructure, provide energy independence and directly improve their security of supply. Community ownership can also facilitate local energy sharing and P2P trading mechanisms, allowing members to buy, sell, and exchange electricity within the community while balancing local supply and demand. Such arrangements have been recognized as important enablers of resilient, decentralized, and consumer-centered energy systems, particularly in regions with limited grid reliability and access to electricity \cite{vanveelen2018community,lage2022energy,wolsink2012social,shared_value}.
Energy communities are also relevant for grids where security of supply is not a concern. In such instances, formation of energy communities facilitates energy autonomy \cite{bonfert2024we, kyriakopoulos2022energy}.

\textit{How does no E6 in CE looks like?}: No facilitation to enable energy communities and no framework for sharing assets.

The six elements of CE are increasingly aligned with policy frameworks, regulatory initiatives, and real-world implementations worldwide. \textbf{Active participation} is being enabled through policies, such as the EU Clean Energy Package, which grants citizens the right to generate, consume, share, and sell renewable energy, transforming consumers into prosumers with bidirectional power flows \cite{european_commission_2026_citizens_package, red_ii_2018, iem_directive_2019}. \textbf{Enhanced choice} is a central focus of the European Commission's \enquote{Unlock Your Power!} campaign, which aims to inform citizens about their energy rights, including selecting suppliers and understanding bills, with only around a third of Europeans currently fully understanding their energy bills \cite{european_commission_2026_unlock}. \textbf{Information and control} through smart meters and digital platforms is being prioritized by regulators like ACER, which calls for wider smart meter adoption to enable real-time demand response, while consumer advocates stress that consumers need tools to see who is accessing their smart meter data to exercise informed consent \cite{acer_2025_unlocking}. \textbf{Automation and controllability} via HEMS is seeing significant growth, with the installed base in Europe and North America reaching 4.5 million in 2024, driven by government incentives and volatile electricity prices \cite{berg_insight_2025_hems}. \textbf{Energy communities and shared assets}, particularly relevant for enhancing security of supply in weak grids, are being facilitated by innovative regulatory measures such as Wallonia's rule allowing prosumers to share surplus solar production within communities, with Italy emerging as a European leader in fostering local energy exchanges through streamlined regulation \cite{interreg_europe_2024_wallonia, concerti_2024_italy, european_parliament_2025_italy, european_commission_2026_recommendation}. Finally, \textbf{self-sufficiency and resilience} through P2P trading and community microgrids are being demonstrated globally, with pilots like Australia's Power Ledger showing potential savings of up to 20\% on bills and increased energy self-sufficiency, while projects in India and Japan are testing blockchain-based platforms for local energy markets \cite{powerledger_2024_australia, powerledger_2025_india_japan, kazam_2025_p2p, repowereu_2022, european_pillar_social_rights}.

\subsection{Proposed definition of consumer empowerment}

The proposed definition of CE considering the six elements detailed in this section is as follows:

\begin{tcolorbox}[colback=green!8!white,colframe=green!75!green]
\textbf{Consumer empowerment} in electricity networks is the ability with which consumers are provided with the access, capabilities, technologies, market opportunities, and institutional arrangements needed to make informed energy decisions, control and automate their energy use, generation and storage assets, participate actively in electricity markets and grid services, collaborate through energy communities, and enhance their energy resilience and self-sufficiency.
\end{tcolorbox}

This proposed definition not only applies to power networks with high levels of reliability but also to weak grids where consumers individually or collectively ensure security of supply.

\pagebreak

\section{Quantification of consumer empowerment}
\label{section3}
In this section, we discuss the dimensions of CE enablers, classify the CE elements discussed in Section \ref{section2} based on enabling dimensions, and finally provide metrics for quantifying CE.


\subsection{Enablers of consumer empowerment}

\begin{figure}[htbp!]
    \centering
    \includegraphics[width=0.63\textwidth]{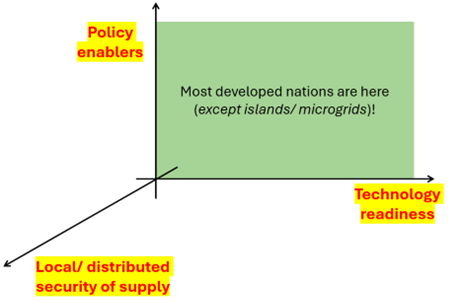} 
    \vspace{-5pt}
    \caption{\small{Dimensions of consumer empowerment.}}
    \label{fig:dimCE}
\end{figure}

The transition towards decentralized and increasingly digitalized electricity systems requires greater participation of electricity consumers in the operation and management of distribution networks. Recent developments, such as FERC Order 2222 in the United States \cite{ferc} and the participation of distributed flexibility in ancillary service markets across Europe \cite{pedro2023opportunities}, demonstrate the growing role of consumers in supporting reliable and efficient power system operation. However, the extent to which consumers can actively participate depends on the availability of several key enablers. As illustrated in Fig.~\ref{fig:dimCE}, three principal dimensions of CE can be identified: technological enablers, policy and regulatory enablers, and the ability to ensure security of supply.

\subsubsection{D1: Availability of Technological Enablers} 
Technological enablers provide the physical and digital infrastructure required for consumers to actively participate in modern electricity systems. These include smart meters, AMI, HEMS, DERs, BESS, controllable appliances, and communication networks. Among these technologies, smart meters play a particularly important role by providing bidirectional communication and near real-time visibility of consumption and generation patterns. Such technologies enable consumers to monitor, control, and optimize their energy usage while supporting DR, flexibility provision, and participation in electricity markets. However, the level of technological readiness varies significantly across countries and regions due to differences in infrastructure investments, digitalization levels, communication technologies, and data accessibility frameworks \cite{cleenwerck2026distribution, kippke2025smart}.

\begin{tcolorbox}[colback=gray!8!white,colframe=blue!75!black]
\small{
\textbf{Smart meter deployment in Italy}: 
Italy was among the first countries to deploy smart meters on a national scale, with Enel's \emph{Telegestore} project installing more than 30 million first-generation smart meters between 2001 and 2006 \cite{enel_telegestore}. Beginning in 2017, the country initiated the rollout of second-generation smart meters (Open Meter project), providing enhanced functionalities such as near real-time consumption information, support for dynamic tariffs, and improved integration of DERs. The widespread availability of smart metering infrastructure has enabled greater consumer access to energy information and constitutes a key technological enabler for CE, facilitating DR, HEMS, and active participation in electricity markets \cite{Giordano_et_al_2011, iea_customer_choice, Depuru2011}.
}
\end{tcolorbox}

\subsubsection{D2: Access to Policy and Regulatory Enablers} 
\label{sec:D2desc}
The availability of technology alone does not guarantee CE. Appropriate policy and regulatory frameworks are required to ensure that consumers can access, utilize, and benefit from available technologies. Regulatory enablers include consumer rights regarding energy data access and ownership, privacy protection mechanisms, dynamic tariff structures, DR programs, flexibility markets, energy-sharing arrangements, and provisions for energy communities. Effective regulatory frameworks create incentives for consumers to provide flexibility services and facilitate coordination among market participants, network operators, aggregators, and consumers. Furthermore, policy frameworks influence the degree to which distributed measurements and flexibility resources can be utilized in centralized, decentralized, or distributed control architectures \cite{7874216}. Consequently, CE depends not only on technological capability but also on the existence of fair and transparent regulatory arrangements that enable active participation \cite{iea_customer_choice,dam2013home,acer_2025_unlocking,eudirective2019}.

\begin{tcolorbox}[colback=gray!8!white,colframe=blue!75!black]
\small{
\textbf{FERC Order No.~2222} (2020) requires Regional Transmission Organizations (RTOs) and Independent System Operators (ISOs) in the United States to remove barriers that prevent aggregated DERs/flexibility from participating in wholesale energy, capacity, and ancillary service markets. By enabling small consumers to aggregate their resources and receive compensation for providing services to the grid, the order enhances CE by expanding market access, increasing consumer choice, and creating new opportunities for active participation in electricity system operation \cite{ferc}.
}
\end{tcolorbox}

\begin{tcolorbox}[colback=gray!8!white,colframe=blue!75!black]
\small{
\textbf{EU Directive 2019/944}, adopted as part of the European Union's Clean Energy for All Europeans Package, establishes common rules for the internal electricity market and formally recognizes active customers, citizen energy communities, demand response, and consumer participation in electricity markets. By granting consumers rights to generate, store, consume, share, and sell electricity, as well as access their energy data and participate in flexibility services, the directive significantly enhances consumer empowerment and supports the transition from passive consumers to active participants in the energy system \cite{eudirective2019}.
}
\end{tcolorbox}

\subsubsection{D3: Ability to Ensure Security of Supply} 
A third dimension of CE is the ability of consumers to maintain or improve their own security of supply. This dimension is particularly relevant for weak-grid systems, remote communities, and regions experiencing frequent outages or load shedding \cite{monergirona2022microgrids,iea_climate_resilience}. In such environments, CE is often driven by the need for energy resilience rather than market participation. Consumers may invest in local generation, battery storage, backup power systems, or community energy resources to reduce dependence on the central grid \cite{monergirona2022microgrids,doe_der}. For example, where outages are infrequent, battery systems can provide continuity of supply during temporary disruptions. In more severe cases, where electricity availability is limited to only a few hours per day, consumers may rely on local microgrids or community-based distribution systems to satisfy their energy needs \cite{monergirona2022microgrids,afdb_energyaccess, AEC2024State}. The ability to maintain electricity supply independently, either individually or collectively through energy communities, therefore represents an important dimension of CE and reflects a consumer's resilience and self-sufficiency in the face of network disturbances \cite{lage2022energy,iea_climate_resilience}.

\begin{tcolorbox}[colback=gray!8!white,colframe=blue!75!black]
\small{
\textbf{Residential solar PV and battery systems in South Africa}:
South Africa provides a recent example of CE driven by security-of-supply concerns. During the severe load-shedding crisis of 2022--2024, many households and commercial consumers invested in rooftop solar PV systems combined with battery storage to maintain electricity supply during grid outages. These investments enabled consumers to reduce dependence on the national grid, improve energy resilience, and enhance self-sufficiency \cite{WATTERSON2025104483, tradegov2023, thango2022, mckay2022, capetimes2024,  mailguardian2024}.
}
\end{tcolorbox}

\subsection{Classifying CE elements in terms of enablers}

The six elements of CE, denoted by E1--E6 (detailed in Sec. \ref{section2}), are closely related to the three enabling dimensions introduced previously, namely the availability of technological enablers (D1), access to policy and regulatory enablers (D2), and the ability to ensure security of supply (D3). While some degree of overlap exists among these dimensions, each CE element is primarily driven by one dominant enabler. Consequently, the elements can be grouped into three categories: E1 and E2 are predominantly technology-driven and therefore align with D1; E3 and E4 rely primarily on regulatory and policy support and are associated with D2; and E5 and E6 focus on resilience, self-sufficiency, and local energy security and therefore align with D3. This classification provides a structured framework for understanding how different enabling conditions support different aspects of CE across diverse electricity systems.

\subsubsection{CE element alignment with D1}

Elements E1 (Facilitating Active Consumer Participation) and E2 (Consumer Enhanced Choices) are primarily associated with the availability of technological enablers, i.e., D1. Active consumer participation requires the presence of enabling technologies such as smart meters, AMI, DERs, communication systems, and digital platforms that allow consumers to interact with energy markets and network operators. Similarly, enhanced consumer choice relies on technological infrastructure that provides access to different suppliers, tariff options, market signals, and energy services. Without the necessary technological capabilities, consumers cannot effectively participate in energy management or exercise meaningful choice. Therefore, E1 and E2 are fundamentally dependent on the availability and accessibility of enabling technologies.

\subsubsection{CE element alignment with D2}

Elements E3 (Access and Control of Consumer Energy Analytics) and E4 (Automation and Controllability) are primarily linked to policy and regulatory enablers. While technology may enable the collection, communication, and processing of energy data, consumers can only access and utilize such information when supported by appropriate regulations governing data ownership, privacy, cybersecurity, and information sharing. Similarly, automated control of DERs through HEMS, aggregators, and flexibility service providers requires regulatory frameworks that permit participation in DR programs, dynamic tariffs, and flexibility markets. Consequently, the benefits of information access, control, and automation depend not only on the availability of technology but also on policies that grant consumers the rights and incentives necessary to use these capabilities effectively.

\subsubsection{CE element alignment with D3}

Elements E5 (Self-Sufficient and Resilient Electricity Consumption) and E6 (Energy Communities and Shared Assets) are primarily associated with the ability to ensure security of supply. Both elements focus on reducing dependence on centralized electricity infrastructure through local generation, storage, community ownership, and shared energy resources. Self-sufficiency and resilience enable consumers to maintain electricity service during network disturbances, while energy communities facilitate collective ownership and management of local energy assets, improving reliability and security of supply at the community level. These attributes are particularly important in weak-grid systems, isolated microgrids, and regions experiencing frequent outages, where CE is often motivated by the need to secure reliable electricity access rather than by participation in competitive energy markets. Therefore, E5 and E6 are naturally classified under the security-of-supply dimension of CE.

\subsection{Metrics for measuring CE}

Quantifying CE is inherently challenging because CE is a multidimensional concept encompassing technology, regulation, market access, autonomy, resilience, and consumer participation. 
Each dimension is normalized within the interval $[0,1]$, where 0 indicates the complete absence of the enabling factor, and 1 represents full availability. Examples of technological enablers include smart meters, advanced metering infrastructure, digital energy platforms, controllable devices, DERs, and HEMS. Policy enablers include access rights to energy data, consumer protection provisions, regulatory support for flexibility markets, dynamic tariffs, P2P trading mechanisms, and energy communities. Security of supply captures the ability of a consumer or community to maintain electricity service through local resources, storage, self-generation, or community-based energy systems.

A straightforward measure of CE can be expressed as \begin{equation} 
CE_1 = D1 \cdot D2 \cdot D3 
\end{equation}
where CE is maximized only when all three dimensions are simultaneously available. This metric reflects the notion that a consumer with access to technology but lacking supporting regulatory frameworks, or vice versa, cannot fully realize the benefits of CE. Similarly, consumers operating in highly reliable grid environments but lacking access to technology or policy support are unable to actively participate in electricity markets and grid services.

However, the multiplicative metric may not adequately represent CE in developing countries or weak-grid environments. In many regions, technological adoption and regulatory maturity remain limited, resulting in low values of $D1$ and $D2$, while consumers often achieve a significant degree of CE through self-sufficiency measures such as diesel backup systems, solar home systems, batteries, microgrids, and local energy communities. In such cases, security of supply may be the dominant objective of CE. A modified metric is therefore proposed: 
\begin{equation} 
CE_2 =  \min(D1 \cdot D2,\; D3) 
\end{equation}

For reliable power networks, this metric becomes: 
\begin{equation} 
CE_2^{\text{strong}} =  D1 \cdot D2.
\end{equation}

For weak power networks, this metric becomes: 
\begin{equation} 
CE_2^{\text{weak}} =  D3. 
\end{equation}

where security of supply acts as an upper bound on realized CE. The metric $CE_2$ reflects the view that technological and policy readiness cannot provide CE beyond the level of service reliability actually experienced by the consumer. Such formulations are particularly relevant in weak-grid systems where resilience and self-sufficiency are often more valuable than sophisticated market participation.

A more general weighted CE index can be defined as \begin{equation} 
CE_w = w_a D1 + w_b D2 + w_c D3 
\end{equation}
subject to  $w_a+w_b+w_c=1$. The weighting factors can be selected according to local priorities. For example, developed electricity systems may assign higher weights to technological and market dimensions, whereas developing systems may prioritize security of supply and resilience. 

In addition to the proposed CE index, several complementary indicators can be used to evaluate specific aspects of CE.

\subsubsection{Self-sufficiency ratio} 
A commonly used metric is the degree to which a consumer satisfies electricity demand using local resources: 
\begin{equation} 
SSR = \frac{E_{\mathrm{local}}}{E_{\mathrm{load}}} 
\end{equation} 
where $E_{\mathrm{local}}$ denotes locally generated and consumed energy and $E_{\mathrm{load}}$ denotes total electricity demand. An $SSR$ approaching unity indicates greater energy autonomy.

\subsubsection{Electricity affordability} 
CE should also translate into improved economic outcomes. An affordability indicator can be defined as 
\begin{equation} 
EA = 1 - \frac{C_{\mathrm{electricity}}^{\text{Active}}}{C_{\mathrm{electricity}}^{\text{Passive}}} 
\end{equation} 
where $C_{\mathrm{electricity}}^{\text{Active}}$ is the annual electricity expenditure for the active consumer and $C_{\mathrm{electricity}}^{\text{Passive}}$ is the cost of electricity for the same consumer without active behaviour.


\subsubsection{Flexibility participation index} 
Consumer participation in flexibility services can be quantified as
\begin{equation} 
FPI = \frac{E_{\mathrm{flex}}}{E_{\mathrm{load}}} 
\end{equation} 
where $E_{\mathrm{flex}}$ represents the energy shifted, curtailed, or controlled in response to market or network signals.

\subsubsection{Energy resilience index} 
Resilience may be quantified as 
\begin{equation} 
ERI = \frac{T_{\mathrm{supported}}}{T_{\mathrm{supported}} + T_{\mathrm{outage}}} 
\end{equation} 
where $T_{\mathrm{supported}}$ represents the duration during which critical loads can be supplied autonomously during a network outage and $T_{\mathrm{outage}}$ denotes the outage duration.

Together, these metrics provide a comprehensive framework for evaluating CE across both advanced and developing electricity systems. While the proposed CE index captures the overall enabling environment, complementary indicators such as self-sufficiency, affordability, flexibility participation, and resilience provide additional insights into the specific benefits realized by consumers.

\pagebreak

\section{Consumer empowerment worldwide}
\label{section4}

Table~\ref{tab:consumer_empowerment_continents} presents a qualitative assessment of the implementation potential of CE elements across different continents. 
The ratings are based on the maturity of electricity markets, digital infrastructure, DER penetration, regulatory support, smart-meter deployment, and consumer participation frameworks reported in the literature \cite{ec_cleanenergy,ceer_smartgrids,doe_gridmodernization,iea_digitalisation,zhang2021smartgrid,idb_smartgrids,irena_latinamerica,afdb_energyaccess,iea_africa,aemo_isp,csiro_der, AEC2024State, IEA2024WEO}.

\textit{\textbf{Disclaimer}: The ratings in Table~\ref{tab:consumer_empowerment_continents} provide very general takeaways, and there may be huge discrepancies among countries and within regions in a particular country.}

\begin{table*}[ht] 
\centering 
\caption{\small{General implementation potential of consumer empowerment elements across continents.}}
\begin{tabular}{|l|c|c|c|c|c|c|} 
\hline 
 & \textbf{E1. }& \textbf{E2. } & \textbf{E3. } & \textbf{E4. } & \textbf{E5. } & \textbf{E6. }  \\ 
\textbf{Continent} & \textbf{Active}& \textbf{Enhanced} & \textbf{Information} & \textbf{Automation} & \textbf{Self-Sufficient} & \textbf{Energy Communities}  \\ 
 & \textbf{Participation} & \textbf{Choice} & \textbf{and Control} & \textbf{Controllability} & \textbf{and Resilient} & \textbf{Shared Assets}  \\ \hline \hline
Europe & \cellcolor{good} & \cellcolor{good} & \cellcolor{good} & \cellcolor{good} & \cellcolor{good} & \cellcolor{good}  \\ \hline 
North America & \cellcolor{good}  & \cellcolor{good} & \cellcolor{good} & \cellcolor{good} & \cellcolor{good} & \cellcolor{moderate} \\ \hline 
South America & \cellcolor{moderate}  & \cellcolor{moderate} & \cellcolor{moderate} & \cellcolor{moderate} & \cellcolor{good} & \cellcolor{good} \\ \hline 
Asia & \cellcolor{moderate} & \cellcolor{moderate} & \cellcolor{good} & \cellcolor{good} & \cellcolor{moderate} & \cellcolor{moderate} \\ \hline 
Africa & \cellcolor{moderate} & \cellcolor{poor} & \cellcolor{poor} & \cellcolor{poor} & \cellcolor{good} & \cellcolor{good} \\ \hline 
Oceania & \cellcolor{good}  & \cellcolor{good} & \cellcolor{good} & \cellcolor{good} & \cellcolor{good} & \cellcolor{good} \\ \hline 
\end{tabular}
\vspace{0.3cm} 
\begin{tabular}{ll} 
\cellcolor{good} & High implementation potential \\ 
\cellcolor{moderate} & Medium implementation potential \\ 
\cellcolor{poor} & Low implementation potential \\ 
\end{tabular} 
\label{tab:consumer_empowerment_continents} 
\end{table*}

Europe demonstrates the highest overall implementation potential across all six elements of CE. Strong market liberalization, widespread smart-meter deployment, ambitious decarbonization policies, and regulatory initiatives such as the European Union's Clean Energy Package have significantly enhanced consumer choice, access to energy data, active participation in electricity markets, and the development of energy communities. Europe is also a global leader in citizen energy cooperatives and community-owned renewable energy projects, enabling both collective ownership and local energy trading \cite{eu_clean_energy,ceer_smartgrid,lage2022}.

North America also exhibits high potential in most CE elements due to advanced digital infrastructure, widespread DR programs, dynamic pricing schemes, and significant deployment of DERs. Active participation and automation are particularly strong in regions with advanced smart-grid initiatives such as California, Texas, and several northeastern states. However, energy communities and collective ownership models are generally less prevalent than in Europe because of differences in regulatory frameworks and electricity market structures \cite{doe_gridmodernization,siano2014}.

Asia presents a heterogeneous landscape. Developed economies such as Japan, South Korea, Singapore, and China have made substantial investments in AMI, automation technologies, and digital platforms, resulting in high implementation potential for information access and controllability. However, the diversity of regulatory environments and market maturity levels across the continent results in moderate scores for active participation, consumer choice, and energy community development \cite{iea_digitalisation,zhang2021smartgrid}.

South America exhibits growing opportunities for CE driven by increasing RES penetration, distributed solar deployment, and rural electrification initiatives. Community-based energy solutions and resilience-oriented microgrids are especially relevant in remote regions. Nevertheless, limited smart-grid deployment and varying levels of electricity market liberalization constrain the broader implementation of advanced consumer-centric services \cite{idb_smartgrids,irena_latinamerica}.

Africa currently faces challenges related to limited grid digitalization, restricted smart-meter deployment, and lower levels of market liberalization. Consequently, information access, automation, and consumer choice remain relatively difficult to implement. However, community-based microgrids, local ownership models, and decentralized renewable energy systems have emerged as effective mechanisms for improving energy access and resilience, particularly in regions with weak transmission and distribution infrastructure \cite{afdb_energyaccess,iea_africa,microgridafrica2022, AEC2024State, IEA2024WEO}.

Oceania, particularly Australia and New Zealand, demonstrates high implementation potential across all six CE elements. High rooftop PV penetration, advanced smart-meter deployment, active retail competition, VPP programs, and strong consumer engagement policies have enabled substantial progress toward consumer-centered electricity systems. The region is frequently cited as a leading example of consumer participation and DER integration at scale \cite{aemo2024,csiro_der}. 

Overall, the results suggest that the implementation of CE is strongly correlated with the maturity of digital infrastructure, regulatory support for market participation, and the availability of DERs. While advanced economies generally perform well across most dimensions, community-based and resilience-oriented empowerment mechanisms can also be highly effective in regions where centralized grid infrastructure remains limited.



\subsection{Real world examples for three dimensions of CE}
In Section \ref{section3}, CE enablers in three dimensions are detailed, i.e., availability of technology enablers, access to policy enablers, and ability to guarantee security of supply. Here we discuss some real-world examples of CE most applicable to one of the three dimensions.  


\subsubsection{Availability of Technology Enablers}
Technology enablers form the backbone of CE in modern energy systems by providing the tools and infrastructure necessary for active participation. Technologies such as smart meters, Internet of Things (IoT) devices, and advanced communication systems allow real-time data exchange between consumers and utilities. This enables better monitoring of energy usage, supports dynamic pricing, and allows consumers to adjust their consumption patterns more efficiently.

Real-world examples highlight the impact of these technologies. In the European Union, countries such as Italy, Spain, the Netherlands, Sweden, and Finland have achieved near-universal adoption of smart meters. These devices enable two-way communication, allowing consumers to access real-time information about their energy usage. 
Sweden completed the large-scale rollout of smart meters in 2009, becoming one of the first countries to mandate monthly meter readings for all customers, thereby improving billing transparency and enabling future demand-response and smart-grid services \cite{ceer_smartgrids, acer2025_data, statista2024_d, europarl2014, researchandmarkets2024, sagemcom2020}.

Similarly, in the United States, utilities have deployed advanced smart grid and DR programs that provide consumers with greater visibility and control over their electricity consumption. Pacific Gas \& Electric (PG\&E) launched its \textit{SmartAC} program in 2007, and by 2013 more than 140{,}000 residential customers had enrolled, providing approximately 165~MW of dispatchable peak-load reduction through the remote cycling of participating air-conditioning units during high-demand events \cite{PGE_SmartAC}. In parallel, utilities such as Consolidated Edison have implemented AMI and ToU tariffs that enable consumers to access detailed consumption information through digital platforms and adjust demand in response to price signals \cite{Faruqui2010,Torriti2012}. These initiatives show how combining AMI technologies, automated DR, and appropriate market mechanisms can enhance CE by improving access to information, enabling automated control, and facilitating active consumer participation in electricity system operation \cite{Faruqui2010,PGE_SmartAC}.

India's Smart Meter National Programme (SMNP), launched in 2017 and implemented by Energy Efficiency Services Limited (EESL), is one of the world's largest smart-meter deployment initiatives. The programme aims to replace approximately 250 million conventional electricity meters with smart prepaid meters to reduce commercial losses, improve billing efficiency, enhance grid operations, and strengthen the financial sustainability of distribution companies \cite{eesl_smnp,goi_rdss}. By 2024, more than 20 million smart meters had been installed across India, providing consumers with near real-time consumption information and prepaid billing functionality, thereby improving consumer awareness, control over electricity expenditure, and participation in a more digitalized electricity ecosystem \cite{eesl_smnp,india_smartmeter2024}.

In Kenya, the deployment of AMI and smart meters by Kenya Power and Lighting Company (a smart grid initiative funded by the World Bank under the Kenya Electricity Modernisation Project) has been used to reduce electricity theft, improve billing accuracy, and enhance network monitoring. The technology provides consumers with improved visibility of their electricity consumption while supporting more efficient utility operations and loss reduction \cite{kenya_power_ami}.

\begin{tcolorbox}[colback=gray!8!white,colframe=blue!75!black]
\small{
In the UK, CE through technology enablers is primarily being pursued through the deployment of smart meters and HEMS, which provide households with real-time visibility, monitoring, and control of their energy consumption. These technologies support demand-side management and load shifting by enabling consumers to respond to ToU tariffs, reduce energy costs, and actively participate in balancing electricity supply and demand. 
A key policy driver is the UK's Smart Systems and Flexibility Plan, which promotes smart energy technologies to help consumers become active participants in the electricity system while contributing to lower energy bills and increased grid flexibility \cite{whittle2020empowering}.
}
\end{tcolorbox}

Together, these examples demonstrate how technology enablers are transforming DNs into more interactive and consumer-centric systems.

\begin{table*}[!t] 
\centering 
\caption{\small{Contribution of smart metering functionalities to the elements of consumer empowerment.}} \label{tab:sm_ce} 
\renewcommand{\arraystretch}{1.3} 
\begin{tabular}{p{4.4cm}cccccc} 
\hline 
\textbf{CE Element} & \makecell{\textbf{Consumption}\\\textbf{Visibility}} & \makecell{\textbf{Near Real-}\\\textbf{Time Data}} & \makecell{\textbf{Price}\\\textbf{Signals}} & \makecell{\textbf{Remote}\\\textbf{Control}} & \makecell{\textbf{DER}\\\textbf{Monitoring}} & \makecell{\textbf{Data Sharing}\\\textbf{\& Market Access}} \\
\hline 
E1: Active Participation & \cellcolor{green!50} & \cellcolor{green!50} & \cellcolor{green!50} & \cellcolor{yellow!40} & \cellcolor{green!50} & \cellcolor{green!50} \\ E2: Enhanced Choice & \cellcolor{green!50} & \cellcolor{yellow!40} & \cellcolor{green!50} & \cellcolor{red!50} & \cellcolor{yellow!40} & \cellcolor{green!50} \\ E3: Information and Control & \cellcolor{green!50} & \cellcolor{green!50} & \cellcolor{yellow!40} & \cellcolor{yellow!40} & \cellcolor{green!50} & \cellcolor{yellow!40} \\ E4: Automation \& Controllability & \cellcolor{green!50} & \cellcolor{green!50} & \cellcolor{green!50} & \cellcolor{yellow!40} & \cellcolor{green!50} & \cellcolor{yellow!40} \\ E5: Self-Sufficiency \& Resilience & \cellcolor{green!50} & \cellcolor{yellow!40} & \cellcolor{yellow!40} & \cellcolor{red!50} & \cellcolor{green!50} & \cellcolor{yellow!40} \\ E6: Energy Communities and Shared Assets & \cellcolor{green!50} & \cellcolor{yellow!40} & \cellcolor{yellow!40} & \cellcolor{red!50} & \cellcolor{green!50} & \cellcolor{green!50} \\ 
\hline 
\end{tabular} 
\end{table*}

Table~\ref{tab:sm_ce} illustrates how different smart metering functionalities contribute to the six elements of CE. Information-related functionalities, such as consumption visibility and near real-time measurements, strongly support information access and consumer awareness. Price signals and data-sharing capabilities facilitate consumer participation in electricity markets and enable informed supplier and tariff choices. Remote control capabilities, when integrated with HEMS and controllable devices, form the basis for automation and flexibility provision. DER monitoring enables consumers to manage local generation and storage resources, thereby increasing self-sufficiency and resilience. Finally, secure data exchange and market access mechanisms are essential for enabling participation in energy communities, peer-to-peer trading, and flexibility markets. The table highlights that no single smart metering functionality enables all empowerment dimensions; rather, CE emerges from the combined availability of multiple complementary technological capabilities.

\subsubsection{Access to Policy Enablers}
Policy enablers establish the regulatory, market, and institutional frameworks that allow consumers to benefit from available technologies and participate actively in the electricity system. Their effectiveness is inherently dependent on the presence of appropriate technological enablers; for example, rights related to data access, dynamic pricing, demand response, or flexibility markets can only be realized when the necessary metering, communication, and control infrastructure is available. Consequently, policy and technological enablers are closely interconnected, with policy frameworks providing the incentives, governance, and access mechanisms needed to unlock the value of consumer-facing technologies and facilitate active participation in decentralized and sustainable distribution networks.

\begin{table*}[!t] 
\centering 
\caption{\small{Examples of CE showcasing policy enabling programs and initiatives worldwide.} }
\label{tab:ce_programs} 
\renewcommand{\arraystretch}{1.25} 
\begin{tabular}{p{1.5cm} p{3.2cm} p{1cm} p{2cm} p{8.45cm}} 
\hline \textbf{Reference} & \textbf{Program Name} & \textbf{Year} & \textbf{Country} & \textbf{Details} \\ \hline 
\cite{eudirective2019} & Directive (EU) 2019/944 & 2019 & European Union & Detail in Sec. \ref{sec:D2desc}\\
\cite{ferc} & FERC Order 2222 & 2020 & United States & Detail in Sec. \ref{sec:D2desc}\\
\cite{acer_2025_unlocking} & Unlocking Flexibility & 2025 & European Union & Regulatory initiative aimed at removing barriers to DR and flexibility services, improving consumer access to flexibility markets and grid support services. \\ 
\cite{PowerLedger2024} & RENeW Nexus & 2024 & Australia & P2P energy trading pilot enabling consumers to exchange locally generated solar energy and increase energy self-sufficiency through blockchain-based trading mechanisms. \\ 
\cite{REC4EU2024} & Energy Sharing Based on Individually Owned Facilities & 2024 & Belgium & Allows prosumers to share surplus solar generation within energy communities, facilitating local energy exchange and collective self-consumption. \\
\cite{Saubhagya2017} & Pradhan Mantri Sahaj Bijli Har Ghar Yojana (Saubhagya) & 2017 & India & National rural electrification programme aimed at providing electricity connections to all willing households. The programme connected more than 28 million households, improving electricity access and enabling future participation in digital energy services. \\
\cite{RDSS2021} & Revamped Distribution Sector Scheme (RDSS) & 2021 & India & Large-scale smart metering programme focused on reducing commercial losses and electricity theft through prepaid smart meters, while improving billing efficiency and giving consumers greater visibility and control over energy consumption. \\
\cite{UKPFR2016} & Enhanced Frequency Response (EFR) Market & 2016 & United Kingdom & National Grid introduced a fast frequency response market that allowed distributed BESS assets, including aggregated consumer-owned resources, to participate in ancillary service provision and receive compensation for grid support. \\
\cite{FCRBelgium2020} & Distributed FCR Participation & 2020 & Belgium & Enabled aggregated residential batteries, EVs, and demand response resources to participate in Frequency Containment Reserve (FCR) markets, allowing small consumers to provide ancillary services traditionally reserved for large generators. \\
\cite{VPPTesla2018} & South Australia VPP & 2018 & Australia & Large-scale VPP aggregating residential rooftop PV and battery systems to improve security of supply, reduce electricity costs, and provide grid services through coordinated operation of consumer-owned assets. \\
\hline 
\end{tabular} 
\end{table*}

Table~\ref{tab:ce_programs} presents a selection of consumer empowerment initiatives implemented across different countries and regulatory contexts. The examples illustrate the diverse pathways through which consumer empowerment can be enabled, including regulatory reforms (e.g., Directive (EU) 2019/944 and FERC Order 2222), community energy and P2P trading initiatives, rural electrification schemes, and mechanisms enabling participation in ancillary service and flexibility markets. Collectively, these programs demonstrate that consumer empowerment extends beyond technology adoption to encompass market access, improved energy access, enhanced system reliability, increased consumer choice, and opportunities for consumers to contribute to electricity system operation and the energy transition actively.

\begin{tcolorbox}[colback=gray!8!white,colframe=blue!75!black]
\small{\textbf{Germany’s Section 14a EnWG}: 
Real-world examples illustrate how such policies empower consumers. In Germany, the Renewable Energy Act (EEG) has been instrumental in promoting the deployment of solar panels and other renewable technologies through feed-in tariffs and financial incentives. This has enabled millions of households to become “prosumers,” actively generating and supplying energy to the grid. Complementing this, Section 14a of the German Energy Industry Act (EnWG) allows grid operators to temporarily reduce the output of controllable energy-consuming devices during network constraints, rather than fully disconnecting them. This applies to larger DERs and ensures that network stability is maintained while still allowing continued operation. Importantly, fairness considerations are emphasized to avoid repeatedly curtailing the same users. Additionally, these policies prevent DSOs from rejecting new connections solely due to capacity limitations, facilitating further DER integration \cite{EEG2023, EEG2000, Jacobsson2006, Hauser2014, EnWG14a2024,  BNetzA14a}.
}
\end{tcolorbox}

\begin{tcolorbox}[colback=gray!8!white,colframe=blue!75!black]
\small{
In the United States, California’s \textbf{Self-Generation Incentive Program} (SGIP) demonstrates another approach to policy support. This program provides financial incentives for installing BESS, enabling consumers to store excess energy—particularly from solar generation—and use it during peak demand periods or outages. Technologies such as residential battery systems have become more widespread due to these incentives, allowing consumers to reduce electricity costs and contribute to grid stability \cite{SGIP2024, KoohiKamali2014, Mongird2020, Barbour2018}.
}
\end{tcolorbox}

At the European level, the General Data Protection Regulation (GDPR) plays an essential role in ensuring that consumer data collected through smart meters and digital devices is securely managed. By safeguarding privacy and establishing clear rules for data usage, GDPR builds consumer trust and encourages participation in energy management and flexibility programs. Together, these policy enablers create a supportive ecosystem where consumers can actively engage in the energy system while ensuring fairness, security, and efficiency.

\subsubsection{Ability to Guarantee Security of Supply}
The ability to guarantee a reliable and secure energy supply is a critical objective in modern DNs, particularly as systems become more decentralized and exposed to external disruptions. With increasing reliance on DERs, consumers are no longer just passive users but can actively enhance supply security through local solutions such as energy storage, microgrids, and self-generation systems. These approaches reduce dependence on centralized infrastructure and improve resilience against outages, natural disasters, and grid constraints.

\begin{tcolorbox}[colback=gray!8!white,colframe=blue!75!black]
\small{
\textbf{CE through Resilient Microgrids in Puerto Rico (2017)}:
Real-world examples demonstrate how local solutions can strengthen security of supply. Following the widespread destruction caused by Hurricane Maria in 2017, several communities in Puerto Rico invested in solar-powered microgrids combined with battery storage to improve energy resilience. One notable example is the community of Adjuntas, where Casa Pueblo developed a solar microgrid capable of supplying critical facilities such as healthcare centers, schools, and community services during extended grid outages. These systems enable communities to maintain essential services even when the main grid is unavailable, significantly improving resilience, energy autonomy, and local security of supply \cite{DOEPR2021, 8644031, AROSVERA2021102055, Chen_et_al_2021, Hirsch2018}.
}
\end{tcolorbox}

\begin{tcolorbox}[colback=gray!8!white,colframe=blue!75!black]
\small{\textbf{Solar Home System Bangladesh}: 
In rural Bangladesh, where grid access has traditionally been limited, solar home systems (SHS) have emerged as a vital solution for ensuring reliable electricity supply. These systems allow households to generate and store their own energy using solar panels and batteries. Large-scale initiatives, such as those led by organizations like Grameen Shakti and supported by IDCOL, have deployed more than six million SHSs, enabling consumers to meet their energy needs without relying on centralized networks \cite{GrameenShakti,IDCOL2023}. Beyond improving energy access, these systems have contributed to greater energy self-sufficiency, improved quality of life, and enhanced resilience in rural communities \cite{Komatsu2011,Mondal2010,Asaduzzaman2013}.
}
\end{tcolorbox}



\begin{tcolorbox}[colback=gray!8!white,colframe=blue!75!black]
\small{\textbf{Brooklyn Microgrid USA}: 
In more advanced urban contexts, projects like the Brooklyn Microgrid in New York illustrate how community-based energy systems can enhance security of supply. This microgrid allows residents to generate, store, and trade locally produced solar energy using digital platforms \cite{Mengelkamp2018,BrooklynMicrogrid}. By enabling P2P energy exchange, the system reduces dependence on the central grid and creates a more resilient and decentralized energy ecosystem \cite{Morstyn2018,Hirsch2018}. Furthermore, such initiatives empower consumers to become active market participants and prosumers, contributing to both local energy resilience and the broader energy transition \cite{ParagSovacool2016}.
}
\end{tcolorbox}


Overall, these examples highlight how empowering consumers with local energy solutions can significantly enhance the security, reliability, and resilience of DNs, especially in the face of growing uncertainties and system challenges.

\subsection{Key take away}
The three dimensions of CE—technology, policy, and security of supply—are essential for creating a sustainable, resilient, and consumer-centric energy system. Real-world examples demonstrate how these dimensions interact to empower consumers, from smart metering in Europe to community microgrids in Puerto Rico and solar home systems in Bangladesh. As technology advances and policies evolve, CE will continue to play a pivotal role in the transition to a decentralized and renewable energy future.



\pagebreak

\section{Expected Outcomes of Consumer Empowerment}
\label{section5}

\begin{figure}[!htbp] 
\centering \begin{tikzpicture}[ every node/.style={align=center,font=\small}, outcome/.style={ draw, rounded corners, fill=blue!10, minimum width=4cm, minimum height=1.5cm }, centerbox/.style={ draw, thick, rounded corners, fill=green!15, minimum width=4cm, minimum height=1.7cm, font=\bfseries }, scale=0.8, transform shape] 
\node[centerbox] (ce) at (0,0) {Consumer\\Empowerment\\ Outcomes}; 
\node[outcome] (soc) at (0,4) { \textbf{Societal \&} \textbf{Environmental}\\\textbf{Outcomes}\\  Renewable/DER adoption\\ Reduced emissions\\ Energy security \\ Decarbonization}; 
\node[outcome] (market) at (5,1.5) { \textbf{Market}\\\textbf{Outcomes}\\  Competition for efficiency\\ Flexibility markets\\ P2P trading }; 
\node[outcome] (cost) at (5,-2.5) { \textbf{Consumer}\\\textbf{Benefits}\\  Lower electricity bills\\ Reduced energy poverty \\
Affordable electricity}; 
\node[outcome] (res) at (-5,-2.5) { \textbf{Energy} \textbf{Autonomy}\\ \textbf{and Resilience}\\  Backup power\\ Self-sufficiency\\ Reliable supply }; 
\node[outcome] (grid) at (-5,1.5) { \textbf{Power System}\\\textbf{Outcomes}\\  Grid stability\\ Congestion relief\\ Deferred upgrades }; 
\draw[->,thick] (ce) -- (soc); \draw[->,thick] (ce) -- (market); \draw[->,thick] (ce) -- (cost); \draw[->,thick] (ce) -- (res); \draw[->,thick] (ce) -- (grid); 
\end{tikzpicture} 
\caption{\small{Expected outcomes of consumer empowerment in electricity networks.} } \label{fig:CE_outcomes} \end{figure}
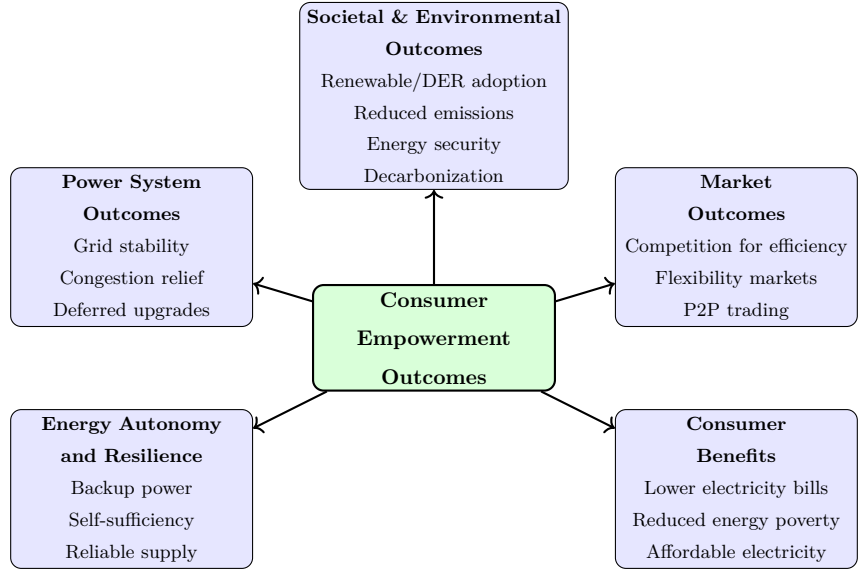

Consumer empowerment is expected to create benefits that extend beyond individual consumers and influence the broader electricity system, energy markets, and society as a whole. As illustrated in Fig.~\ref{fig:CE_outcomes}, the outcomes of consumer empowerment can be grouped into five interrelated categories: power system outcomes, market outcomes, consumer benefits, energy autonomy and resilience, and societal and environmental outcomes. Together, these outcomes demonstrate how enabling consumers to actively manage, generate, store, and exchange electricity can contribute to a more reliable, efficient, affordable, resilient, and sustainable energy system. The following subsections discuss each of these outcome categories in greater detail.

\subsection{Power System Outcomes} 
CE can significantly improve the operation and planning of electricity networks. Through demand response, demand-side flexibility, distributed storage, and coordinated operation of DERs, empowered consumers can help reduce peak demand, alleviate network congestion, and improve voltage and frequency stability \cite{iea_customer_choice,Burger2019}. Active participation of consumers allows DSOs and TSOs to utilize flexible demand and distributed resources as alternatives to conventional network reinforcement, thereby deferring costly investments in transmission and distribution infrastructure \cite{acer_2025_unlocking}. As a result, CE contributes to improved grid stability, more efficient asset utilization, and increased hosting capacity for RES \cite{HAFEZIMAGHAM2025101894}.

\subsection{Market Outcomes} 
CE facilitates the development of more competitive and efficient electricity markets by enabling consumers to actively participate in energy trading and flexibility services. Through technologies such as smart meters, aggregators, and digital trading platforms, consumers can respond to price signals, participate in demand response programs, and provide ancillary services using distributed energy resources \cite{ferc,ParagSovacool2016}. Emerging market models, including P2P energy trading, local flexibility markets, and VPPs, create additional revenue streams for consumers while improving market efficiency and competition \cite{Mengelkamp2018,Morstyn2018}. These developments help align consumer behavior with system needs, leading to more efficient utilization of energy resources.

\subsection{Consumer Benefits} 
One of the primary objectives of CE is to improve consumer welfare. Access to real-time information, dynamic tariffs, distributed generation, and energy management technologies allows consumers to optimize electricity consumption and reduce energy costs \cite{Faruqui2010,europaProtectingEmpowering}. Participation in flexibility markets and demand response programs can further provide direct financial compensation for supporting grid operation \cite{Burger2019,iea_customer_choice}. By increasing access to affordable electricity services and enabling greater control over energy expenditures, CE can also contribute to reducing energy poverty\footnote{Energy poverty is a complex and multidimensional problem, which describes a situation where a household lacks access to essential energy services—such as heating, cooling, lighting, and powering appliances—needed for a decent standard of living and health \cite{JRC_News_2024, lage2022energy}.} and enhancing overall consumer satisfaction.

As already mentioned, CE enables consumers to optimize energy consumption, utilize self-generated electricity, participate in flexibility markets, and select tariff structures that better align with their needs. In addition, policy initiatives such as the European Commission's \emph{Affordable Energy Action Plan} aim to lower electricity bills by reducing electricity taxation, removing non-energy costs from bills, improving network charge efficiency, and enabling consumers to switch suppliers and access affordable renewable energy \cite{europaAffordableEnergy}. Lower energy costs can further contribute to addressing energy poverty, which remains a major challenge worldwide. Within Europe alone, more than 46 million people are estimated to be affected by energy poverty, which occurs when households are unable to secure adequate energy services without compromising their health and well-being \cite{europaEnergyPoverty}.

\begin{tcolorbox}[colback=gray!8!white,colframe=blue!75!black]
\small{
\textbf{CE for affordability:} Real-world experience from Denmark demonstrates that empowering local actors and consumers through coordinated energy planning can significantly lower energy costs. By aggregating demand, sharing infrastructure costs, and enabling consumers to benefit from communal energy solutions, Danish district heating systems provide heating costs that are lower than many individual heating alternatives while simultaneously reducing risk, improving reliability, and increasing long-term affordability for households \cite{nilsson2024empowered}.
}
\end{tcolorbox}

\subsection{Energy Autonomy and Resilience} 
\label{sec:energyAutonomy}
CE enhances resilience by enabling consumers and communities to improve their self-sufficiency and maintain electricity supply during disruptions. DG, BESS, microgrids, and energy communities can reduce dependence on centralized electricity infrastructure and provide backup power during outages \cite{Hirsch2018}. In regions with weak or unreliable grids, these capabilities are particularly valuable for ensuring security of supply and maintaining critical energy services \cite{Monyei2018,Komatsu2011}. Consequently, empowered consumers are better positioned to withstand extreme weather events, infrastructure failures, and other disruptions while improving local energy autonomy.

\subsection{Societal and Environmental Outcomes} 
CE supports broader societal and environmental objectives by accelerating the adoption of renewable energy technologies and enabling progress toward decarbonization goals. Empowered consumers are more likely to invest in rooftop PV, BESS, EVs, and energy-efficient technologies when provided with appropriate incentives and information \cite{ParagSovacool2016,Jacobsson2006}. 
Further, as consumers with DERs become active participants in the energy system, they are more likely to invest in clean energy technologies and adopt energy-efficient practices \cite{jrc_engagement}.
Increased participation in renewable energy generation and flexibility services facilitates higher integration of clean energy resources while reducing greenhouse gas emissions \cite{iea_customer_choice,ConsumersInternational2023, p2p_aggregation}. 
CE is financially crucial towards coping with the rapid energy transition, as showcased in Fig. \ref{fig:approached}.

CE also facilitates the emergence of new business models and market structures. The increasing participation of consumers in electricity markets has stimulated the development of aggregators, VPPs, P2P energy trading platforms, local flexibility markets, and energy communities. These innovations create new economic opportunities while helping system operators access distributed flexibility resources in a more efficient manner.

Overall, CE contributes to a more resilient electricity system through enhanced local flexibility, reduced peak demand, and lower renewable integration costs. It can lower energy bills through optimized energy consumption and market participation, while simultaneously supporting decarbonization objectives through greater adoption of clean energy technologies and more sustainable consumption patterns. Thus, CE is a key enabler of the transition toward reliable, affordable, and carbon-neutral electricity systems.

\pagebreak

\section{Challenges of Consumer Empowerment} 
\label{section6}

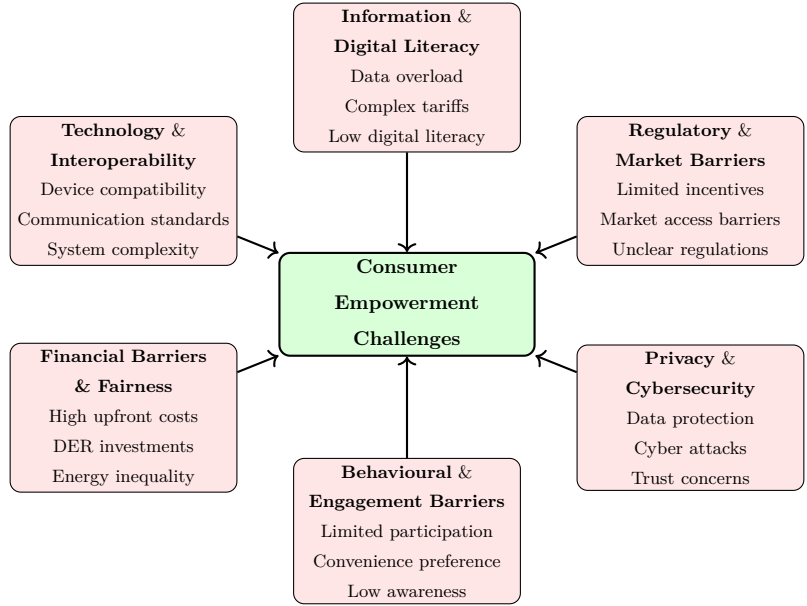
\begin{figure*}[!htbp] 
\centering \begin{tikzpicture}[ node distance=2cm, every node/.style={font=\small, align=center}, challenge/.style={ draw, rounded corners, fill=red!10, minimum width=4cm, minimum height=1.6cm }, centerbox/.style={ draw, thick, rounded corners, fill=green!15, minimum width=4.5cm, minimum height=1.8cm, font=\bfseries }, scale=0.75, transform shape ] 
\node[centerbox] (CE) at (0,0) {Consumer\\Empowerment\\Challenges}; 
\node[challenge] (Info) at (0,4) { \textbf{Information} \&\\ \textbf{Digital Literacy}\\  Data overload\\ Complex tariffs\\ Low digital literacy }; \node[challenge] (Tech) at (-5,2) { \textbf{Technology} \&\\ \textbf{Interoperability}\\  Device compatibility\\ Communication standards\\ System complexity }; \node[challenge] (Reg) at (5,2) { \textbf{Regulatory} \&\\ \textbf{Market Barriers}\\  Limited incentives\\ Market access barriers\\ Unclear regulations }; 
\node[challenge] (Finance) at (-5,-2) { \textbf{Financial Barriers}\\ \textbf{\& Fairness}\\  High upfront costs\\ DER investments\\ Energy inequality }; \node[challenge] (Privacy) at (5,-2) { \textbf{Privacy }\&\\ \textbf{Cybersecurity}\\  Data protection\\ Cyber attacks\\ Trust concerns }; \node[challenge] (Behaviour) at (0,-4) { \textbf{Behavioural} \&\\ \textbf{Engagement Barriers}\\  Limited participation\\ Convenience preference\\ Low awareness }; 
\draw[thick,->] (Info) -- (CE); \draw[thick,->] (Tech) -- (CE); \draw[thick,->] (Reg) -- (CE); \draw[thick,->] (Finance) -- (CE); \draw[thick,->] (Privacy) -- (CE); \draw[thick,->] (Behaviour) -- (CE); \end{tikzpicture} \caption{\small{Major challenges affecting the realization of CE in modern electricity networks. Successful CE requires addressing information, financial, technological, regulatory, cybersecurity, and behavioural barriers simultaneously.} } \label{fig:consumer_empowerment_challenges} \end{figure*}

Despite its potential benefits detailed in previous sections, CE faces several technical, economic, regulatory, and social challenges that may limit its adoption and effectiveness. While modern electricity systems increasingly rely on active consumer participation, many consumers lack the resources, knowledge, technologies, or incentives required to engage effectively in energy markets and grid-support activities \cite{ci_empowerment,enel_ci}. 
As illustrated in Fig.~\ref{fig:consumer_empowerment_challenges}, the realization of CE depends not only on the availability of enabling technologies and supportive policies but also on the ability of consumers to effectively access, understand, trust, and utilize these opportunities. The challenges can be broadly classified into six categories: information and digital literacy, regulatory and market barriers, privacy and cybersecurity concerns, behavioural and engagement barriers, financial barriers, and technology and interoperability challenges. Addressing these barriers in a coordinated manner is essential for ensuring inclusive, equitable, and sustainable consumer participation in future electricity systems.

\subsection{Information and Digital Literacy Challenges} 
A major challenge is the limited availability and accessibility of information.
However, access to information alone does not guarantee CE. Modern electricity systems generate large volumes of data through smart meters, digital platforms, and energy management systems, which can overwhelm consumers and hinder effective decision-making. Furthermore, many consumers lack the digital literacy and technical knowledge required to interpret energy data, compare tariff structures, or participate in flexibility programs \cite{ConsumerPolicy2018,ConsumersInternational2023}. Complex tariff structures and limited understanding of electricity markets can reduce participation and prevent consumers from fully benefiting from available technologies and services. Consequently, consumer education, intuitive interfaces, and simplified information presentation remain critical requirements for successful empowerment.

\subsection{Regulatory and Market Barriers} 
Regulatory and market frameworks play a fundamental role in determining the extent to which consumers can participate in electricity markets. In many jurisdictions, consumers face barriers such as unclear regulations, limited access to flexibility markets, insufficient incentives, and market designs that favor large participants over distributed energy resources \cite{ferc,acer_2025_unlocking,ParagSovacool2016}. Regulatory uncertainty can also discourage investments in distributed generation, storage, and demand response technologies. Therefore, effective CE requires regulatory frameworks that facilitate fair participation, transparent market access, and compensation mechanisms aligned with the services provided by consumers.

\begin{tcolorbox}[colback=orange!8!white,colframe=orange!75!orange]
\small{
\textbf{Complexity and low switching}: consumers often prefer simple tariff structures over dynamic pricing, and switching (tariffs, suppliers, etc.)-- a key indicator used by regulators -- has remained comparatively low even where information is provided \cite{he2013engage, ConsumerPolicy2018}.
}
\end{tcolorbox}

\subsection{Privacy and Cybersecurity Challenges} 
CE increasingly relies on the collection, exchange, and processing of granular energy consumption data. While such information is essential for advanced energy management and flexibility services, it also raises significant privacy and cybersecurity concerns \cite{McKenna2012,Depuru2011}. Consumers may be reluctant to share detailed energy data if they fear misuse of personal information, unauthorized access, or cyberattacks targeting connected devices and digital platforms. The growing deployment of smart meters, HEMS, and IoT devices further expands the potential attack surface. Establishing robust cybersecurity measures, transparent data governance policies, and consumer trust is therefore essential for enabling widespread participation.

\begin{tcolorbox}[colback=orange!8!white,colframe=orange!75!orange]
\small{
\textbf{Technical and cyber risks}: broader smart grid challenges, including data collection and management, cybersecurity, and consumer acceptance, remain open issues affecting the practical realisation of CE \cite{shaukat2018survey}.
}
\end{tcolorbox}

\subsection{Behavioural and Engagement Barriers} 
CE ultimately depends on consumer engagement. However, many consumers exhibit limited interest in actively managing their electricity consumption, even when economic incentives are available \cite{he2013engage,Faruqui2010}. Behavioral studies show that consumers often prioritize convenience and simplicity over active participation in complex energy programs. Limited awareness of available technologies, low trust in energy providers, and uncertainty regarding potential benefits can further reduce participation rates. As a result, successful CE requires not only technological and regulatory solutions but also consumer-centric program design, effective communication, and automated participation mechanisms that minimize effort while maximizing benefits.

\begin{tcolorbox}[colback=orange!8!white,colframe=orange!75!orange]
\small{
\textbf{Trust and behavioural uncertainty}: CE can potentially be hindered by a lack of trust and by heterogeneous motivational factors, ranging from bill reduction to comfort and environmental concern \cite{jrc_engagement, nilsson2024empowered}.
}
\end{tcolorbox}

CE increasingly depends on the collection, exchange, and processing of detailed energy consumption data. While data-driven energy management offers substantial benefits, consumers may be reluctant to participate if they perceive risks related to privacy breaches, unauthorized data access, or cyberattacks on connected devices and energy management systems. Building consumer trust through transparent governance, data protection measures, and cybersecurity standards is therefore essential for successful CE initiatives \cite{europaProtectingEmpowering,oecd_consumers}.

\subsection{Financial Barriers \& Fairenss} 

Many CE technologies, including rooftop PV systems, BESS, EVs, and HEMS, require significant upfront investments that remain unaffordable for many households \cite{ParagSovacool2016,Burger2019}. Consequently, the benefits of CE, such as lower electricity costs, participation in flexibility markets, enhanced resilience, and self-consumption, may accrue disproportionately to consumers with greater financial resources and access to enabling technologies. In contrast, vulnerable consumers may be excluded from these opportunities while continuing to bear a share of network and policy costs, potentially exacerbating existing energy inequalities. Addressing these challenges requires inclusive policy frameworks, targeted support measures, and innovative financing mechanisms to ensure equitable access to CE across different consumer groups.

\begin{tcolorbox}[colback=orange!8!white,colframe=orange!75!orange]
\small{
\textbf{Vulnerable and low-income consumers}: 
The CE paradigm alone may be insufficient to address the needs of vulnerable and low-income consumers, as participation often depends on access to financial resources, technology, and digital capabilities. Consequently, the benefits of CE may not be distributed equitably across society, motivating a shift from a narrow focus on individual empowerment toward a broader energy ecosystem based on shared value, inclusiveness, and equitable access to energy services \cite{shared_value}.
}
\end{tcolorbox}

\subsection{Technology and Interoperability Challenges} 
The effectiveness of consumer empowerment depends on the ability of diverse technologies and stakeholders to communicate and operate seamlessly. Consumers increasingly interact with smart meters, DERs, batteries, EVs, home energy management systems, aggregators, and market platforms, often supplied by different vendors. A lack of interoperability, inconsistent communication standards, and proprietary technologies can limit functionality, increase costs, and create integration challenges \cite{shaukat2018survey,Hirsch2018}. Moreover, managing multiple interconnected devices increases system complexity and can discourage adoption among less technically inclined consumers. Continued standardization efforts and open communication protocols are therefore essential for enabling scalable and interoperable consumer participation in future electricity systems.

Finally, consumer engagement cannot be assumed. Numerous studies have shown that many consumers prioritize convenience and reliability over active market participation. Even when financial incentives exist, consumers may be unwilling to continuously monitor prices, adjust consumption patterns, or manage DERs. Therefore, successful empowerment strategies increasingly rely on automation, smart controls, aggregators, and consumer-centric market designs that reduce participation complexity while preserving consumer choice and autonomy \cite{iea_customer_choice,ci_empowerment}.

Overall, CE requires more than technology deployment alone. Its success depends on addressing economic, regulatory, behavioral, technological, and social barriers simultaneously. Without appropriate consumer protections, inclusive policies, and supportive market frameworks, empowerment initiatives risk creating unequal access to benefits and limiting the contribution that consumers can make to the operation and transition of future electricity systems.


\pagebreak

\section{Conclusions \& Future works}
\label{section7}

Consumer empowerment (CE) is emerging as a fundamental pillar of modern electricity systems and a critical enabler of the energy transition. The increasing penetration of distributed energy resources, electrification of transport and heating, digitalization, and the growing need for flexibility are transforming electricity consumers from passive users into active participants capable of supporting power system operation and decarbonization objectives. 

This paper reviewed the need for CE and proposed a holistic definition encompassing six elements: active participation, enhanced choice, information and control, automation and controllability, self-sufficiency and resilience, and energy communities with shared ownership. Together, these dimensions capture the evolving role of consumers in future electricity systems and provide a common framework for researchers, policy makers, regulators, and network operators. 

Recognizing that CE is a multidimensional concept, the paper proposed a framework for its quantification based on technology availability, policy enablement, and security-of-supply considerations. Complementary indicators addressing self-sufficiency, affordability, flexibility participation, and resilience were also presented to facilitate a more comprehensive assessment of CE outcomes across different electricity systems.

The analysis demonstrates that CE can deliver significant benefits, including lower electricity bills, increased consumer choice, improved integration of distributed energy resources, enhanced network flexibility and resilience, deferred network investments, greater energy autonomy, and the development of innovative business models such as aggregators, flexibility markets, and peer-to-peer energy trading. However, substantial challenges remain, including information asymmetries, digital literacy gaps, high technology costs, regulatory and market barriers, interoperability issues, cybersecurity concerns, and unequal access to CE opportunities. 
Overall, CE should not be viewed solely as a consumer-rights initiative but as a strategic mechanism for improving the efficiency, affordability, reliability, resilience, and sustainability of future electricity networks. 

Future work should focus on developing standardized CE metrics, evaluating real-world implementations, and designing consumer-centric regulatory frameworks that ensure equitable access to the benefits of participation in evolving electricity markets.

\pagebreak

\bibliographystyle{IEEEtran}
\bibliography{reference2.bib} 

\end{document}